\documentclass[useAMS,usenatbib]{mn2e}

\usepackage{epsfig,color}
\usepackage{natbib}
\bibpunct{(}{)}{;}{a}{}{,}
\def\etal{{\rm et al.}}
\def\eg{{\rm e.g.}}

\def\cf{{\rm cf.}}

\def\spose#1{\hbox to 0pt{#1\hss}}
\def\ltsimm{\mathrel{\spose{\lower 3pt\hbox{$\sim$}}
        \raise 2.0pt\hbox{$<$}}}
\def\gtsimm{\mathrel{\spose{\lower 3pt\hbox{$\sim$}}
        \raise 2.0pt\hbox{$>$}}}
\def\Mdot{\hbox{${\dot M}$}}

\def\km{{\rm\thinspace km}}
\def\cm{{\rm\thinspace cm}}

\def\s{{\rm\thinspace s}}
\def\yr{{\rm\thinspace yr}}
\def\g{{\rm\thinspace g}}
\def\kmps{\hbox{${\rm\km\s^{-1}\,}$}}

\def\erg{{\rm\thinspace erg}}
\def\eV{{\rm\thinspace eV}}
\def\GeV{{\rm\thinspace GeV}}
\def\Hz{{\rm\thinspace Hz}}

\def\ster{{\rm\thinspace ster}}
\def\ergps{\hbox{${\rm\erg\s^{-1}\,}$}}

\def\Msol{\hbox{${\rm\thinspace M_{\odot}}$}}
\def\Lsol{\hbox{${\rm\thinspace L_{\odot}}$}}
\def\Msolpyr{\hbox{${\rm\Msol\yr^{-1}\,}$}}
\def\pcm{\hbox{${\rm\cm^{-1}\,}$}}
\def\pcm2{\hbox{${\rm\cm^{-2}\,}$}}
\def\pcm3{\hbox{${\rm\cm^{-3}\,}$}}
\def\ergpscm3Hz{\hbox{${\rm\ergps\cm^{-3}\Hz^{-1}\,}$}}
\def\ergpscm3Hzster{\hbox{${\rm\ergps\cm^{-3}\Hz^{-1}\ster^{-1}\,}$}}
\def\gpcm3{\hbox{${\rm\g\cm^{-3}\,}$}}
\def\ergpcm2{\hbox{${\rm\erg\cm^{-2}\,}$}}
\def\ergpcm3{\hbox{${\rm\erg\cm^{-3}\,}$}}
\def\phpscm2{\hbox{${\rm photons\s^{-1}\cm^{-2}\,}$}}
\def\wr{{\rm WR\thinspace}}

\title[Models of the Colliding Winds Binary WR\thinspace140]{Radio, X-ray, and $\gamma$-ray Emission Models of the Colliding Winds Binary WR\thinspace140}
\author[J.~M.~Pittard and S.~M.~Dougherty]
{J. M. Pittard$^{1}$\thanks{E-mail: jmp@ast.leeds.ac.uk} and
S.~M.~Dougherty$^{2}$\\
$^{1}$School of Physics and Astronomy, The University of
        Leeds, Woodhouse Lane, Leeds LS2 9JT, UK\\
$^{2}$National Research Council of Canada, Herzberg Institute for
        Astrophysics, Dominion Radio Astrophysical Observatory, \\
        P.O. Box 248, Penticton, British Columbia V2A 6J9, Canada}

\begin{document}

\date{Accepted ... Received ...; in original form ...}

\pagerange{\pageref{firstpage}--\pageref{lastpage}} \pubyear{2005}

\maketitle

\label{firstpage}

\begin{abstract}
We use hydrodynamical models of the wind-collision region (WCR)
in the archetype colliding-wind system WR\thinspace140 to determine
the spatial and spectral distribution of the radio, X-ray and
$\gamma$-ray emission from shock accelerated electrons. Our
calculations are for orbital phase 0.837 when the observed radio
emission is close to maximum. Using the observed thermal X-ray
emission at this phase in conjunction with the radio emission to
constrain the mass-loss rates, we find that the O-star mass-loss rate
is consistent with the reduced estimates for O4-5 supergiants by
\citet{Fullerton:2006}, and the wind momentum ratio, $\eta = 0.02$.
This is independent of the opening angle deduced from
radio VLBI observations of the WCR that we demonstrate fail to
constrain the opening angle.

We show that the turnover at $\sim3$~GHz in the radio emission is due
to free-free absorption, since models based on the Razin effect have
an unacceptably large fraction of energy in non-thermal electrons. We
find the spectral index of the non-thermal electron energy
distribution is flatter than the canonical value for diffusive shock
acceleration (DSA), namely $p<2$. Several mechanisms are
discussed that could lead to such an index. Our inability to obtain
fits to the radio data with $p>2$ does not exclude the possibility of
shock modification, but stronger evidence than currently exists is
necessary for its support.

Tighter constraints on $p$ and the nature of the shocks in
WR\thinspace140 will be obtained from future observations at MeV and
GeV energies, for which we generally predict lower fluxes than
previous work. Since the high stellar photon fluxes prevent the
acceleration of electrons beyond $\gamma \gtsimm 10^{5}-10^{6}$, TeV
emission from colliding-wind binary (CWB) systems will provide
unambiguous evidence of pion-decay emission from accelerated ions. We
finish by commenting on the emission and physics of the multiple wind
collisions in dense stellar clusters, paying particular attention to
the Galactic Center.
\end{abstract}


\section{Introduction}
\label{sec:intro}
Massive binary systems containing a Wolf-Rayet (WR) and O-type star
often exhibit synchrotron emission arising from relativistic electrons
in the presence of a magnetic field \citep{Dougherty:2000b}. The
non-thermal electrons are widely thought to be accelerated through
diffusive shock accleration (DSA), though other mechanisms are also
possible \citep[\eg,][]{Jokipii:1987,Jardine:1996}.
High-spatial-resolution radio observations
\citep[\eg,][]{Williams:1997,Dougherty:2000,Dougherty:2005} have
revealed that the acceleration site is the region where the massive
stellar winds collide - the wind-collision region
(WCR). Thus, colliding-wind binary (CWB) systems present an important
laboratory for investigating the underlying physics of particle
acceleration since they provide access to higher mass, radiation and
magnetic field energy densities than in supernova remnants (SNRs),
which have been widely used for such work. 

CWB systems are also ideal for understanding related, but more
complicated, systems involving colliding winds, such as the cluster of
massive stars in the central parsec of the Galaxy \citep{Ghez:2005},
and young, dense stellar clusters (super star clusters), such as the
Arches \citep{Figer:1999}, NGC~3603 \citep{Moffat:2002}, and R136
\citep{Figer:1999b}.  Non-thermal radio emission has recently been
detected from the Arches cluster \citep{Yusef-Zadeh:2003}, and the
non-thermal radio filaments and ``streaks'' near the Galactic Center
may have their origin in CWBs and cluster winds
\citep{Yusef-Zadeh:2003b}.

In this paper we model the thermal and non-thermal radio, X-ray
and $\gamma$-ray emission from one of the most extensively observed
CWB systems, WR\thinspace140. In Sec.~\ref{sec:ntpart_spec} we
describe our assumptions for the non-thermal particle spectrum.
Sec.~\ref{sec:gamma} contains details of our implementation of the
various processes responsible for the non-thermal X-ray and
$\gamma$-ray emission. Details of the modelling of the radio emission,
including several cooling processes, and various emission and
absorption mechanisms, can be found in \citet{Dougherty:2003} and
\citet{Pittard:2006}. The application of our model to observations of
WR\thinspace140 is described in Sec.~\ref{sec:wr140}, and the
implications of our modelling are discussed in Sec.~\ref{sec:discuss}.
In Sec.~\ref{sec:summary}, we summarize and note future directions.

\section{The non-thermal particle spectrum and magnetic field}
\label{sec:ntpart_spec}
In this work, it is assumed that the non-thermal particle distribution
can be approximated as a power-law $n(\gamma)\propto \gamma^{-p}$,
where $\gamma$ is the Lorentz factor and $p$ is a free-parameter
determined from model fits (primarily) to the radio emission.  A
number of mechanisms can produce such a distribution, as we now
discuss along with our assumptions concerning the normalization of the
non-thermal particle spectrum and the magnetic field within the WCR.

\subsection{Acceleration mechanisms and spectral index of non-thermal 
particles} 
There are several possible mechanisms for accelerating particles
in CWB systems, including DSA, reconnection, and many turbulent
processes.  The spectral index of the non-thermal particles can, in
each instance, span a wide range of possibilities, as discussed
below. An estimate of which mechanism is likely to be dominant can be
made by comparing their relative efficiencies.

\subsubsection{Reconnection}
In CWB systems, particles may be accelerated at the contact
discontinuity between the two winds as magnetic field lines from the
two stars are forced together and reconnect \citep{Jardine:1996},
and/or within the volume of the WCR as tangled field lines
reconnect. Reconnection can occur either as a quasi-steady, or as an
explosive, process. The energy released is transferred into thermal
energy through direct Joule heating of the plasma, into the kinetic
energy of macroscopic motions, and into accelerated particles.
Unfortunately, the role of magnetic reconnection in accelerating
particles is complex, not least because of the many different types of
reconnection, and the intricate magnetic environment that is usually
associated with reconnection. This environment can accelerate
particles directly (through strong electric fields), stochastically
(i.e. through the second-order Fermi process due to high turbulence),
and at the MHD shock waves which are an integral part of a
reconnecting field \citep[i.e. through DSA - see][and references
therein]{Priest:2000}.  Reconnection is usually thought to produce
steep spectra i.e. $p>2$, such as is seen in solar flares.

A rough upper limit of the power available for particle acceleration
due to reconnection can be estimated as follows. If reconnection
occurs as a quasi-steady process, the power released in a current
sheet of area $A$ is
\begin{equation}
\label{eq:recon}
P \approx \frac{\pi}{4\;{\rm ln}\;R_{\rm m}}\frac{B^{2}}{8 \pi}v_{\rm A} A,
\end{equation}
where $R_{\rm m}$ is the magnetic Reynold's number, $B$ is the
magnetic field strength, and $v_{\rm A}$ is the Alfv\'{e}n speed
\citep{Petschek:1964}.  If reconnection occurs at the contact
discontinuity, the area of the current sheet, $A$, is approximately
the area of the apex of the WCR i.e. $A\sim \pi r_{\rm O}^{2}$,
where $r_{\rm O}=\eta^{1/2}D/(1+\eta^{1/2})$ is the distance from the
stagnation point of the WCR to the center of the O star, $D$ is the
separation of the stars, and the wind momentum ratio $\eta=\Mdot_{\rm
O}v_{\rm O}/\Mdot_{\rm WR}v_{\rm WR}$ where $\Mdot_{\rm O}$ and
$\Mdot_{\rm WR}$ are the mass-loss rates of the O and WR stars with
terminal wind speeds of $v_{\rm O}$ and $v_{\rm WR}$ respectively.
For our model~B (see Table~\ref{tab:wr14067_params}),
$R_{\rm m}\sim10^{20}$, $A\sim5\times10^{27}\;{\rm cm^{2}}$, $v_{\rm
A} \approx 6\times10^{7} \;{\rm cm}\;{\rm s^{-1}}$, and $B=1.2$~G, so
$P\sim3\times10^{32}\ergps$. Since $P$ is strongly dependent on $B$,
the available power from reconnection could be considerably higher
than this estimate, and may be further augmented by reconnection
occuring throughout the WCR region. While we expect that only a
fraction of this total power will end up in non-thermal particles,
there is probably sufficient reconnection luminosity to produce the
synchrotron emission observed from WR\thinspace140 ($\sim5 \times 10^{32} 
\;\ergps$). However, it may struggle to supply the energy in 
IC $\gamma$-ray emission, which is expected to be $\sim 10\times$ 
greater (see Table~\ref{tab:nt_lum_models}).

\subsubsection{Turbulent mechanisms}
Particles may also be accelerated stochastically, by magnetic
scattering off turbulent motions \citep{Fermi:1949}. In this
mechanism, non-thermal particles either gain or lose energy depending
on whether the magnetic field is moving towards or away from the
particle. However, as the scattering probability is proportional to the
relative velocity between the particle and the plasma, there is a
slight excess of head-on scatterings and a mean increase in
momentum. This process is known as the second-order Fermi mechanism, and in
principle, hard power-law spectra ($p<2$) may be obtained. 
The second-order Fermi mechanism is slower and less efficient than
DSA (a first-order Fermi process) in our models of WR\thinspace140, 
since the pre-shock Alfv\'{e}nic
Mach number, $M_{\rm A}=v\sqrt{4 \pi \rho}/B\gtsimm3$ - see
Table~\ref{tab:wr14067_params}.   Another possibile mechanism is 
acceleration in turbulence from resistive magnetohydrodynamical
fluctuations \citep{Nodes:2004} that can produce hard power-law
spectra with $p<2$.

\subsubsection{Diffusive shock acceleration}
\label{sec:dsa}
DSA is a variant of the original Fermic mechanism, in which particles
are accelerated by scattering back and forth across a shock. The key
difference is that fast particles gain energy {\em each} time they are
scattered back across the shock, until their eventual loss downstream
of the shock. 
To a first approximation, DSA produces a power-law
energy distribution of non-thermal particles \citep[see][for a
discussion of departures from this]{Pittard:2006}. 
The maximum power that is available to DSA is the kinetic luminosity
of that part of the stellar winds which are processed through the WCR
(in model~B this is $\sim10^{36} \ergps$). Therefore, DSA easily meets the
energy requirements for producing the synchrotron emission seen from
WR\thinspace140 ($\sim5 \times 10^{32} \;\ergps$).

At a single, parallel shock the value of $p$ is determined by the
compression ratio of the scattering centers across the shock, $r_{\rm
k}$ \citep[e.g.,][]{Bell:1978}.  In the strong shock limit, the gas
compression ratio $r=4$; if $r_{\rm k} \approx r$, then
$p\approx2$. However, $p$ can be greater or smaller than this value
depending on the values of $r$ and the ratio of gas to magnetic
pressure in the upstream flow, which together determine $r_{\rm k}$
\citep{Schlickeiser:2002}.

\begin{figure}
\begin{center}
\psfig{figure=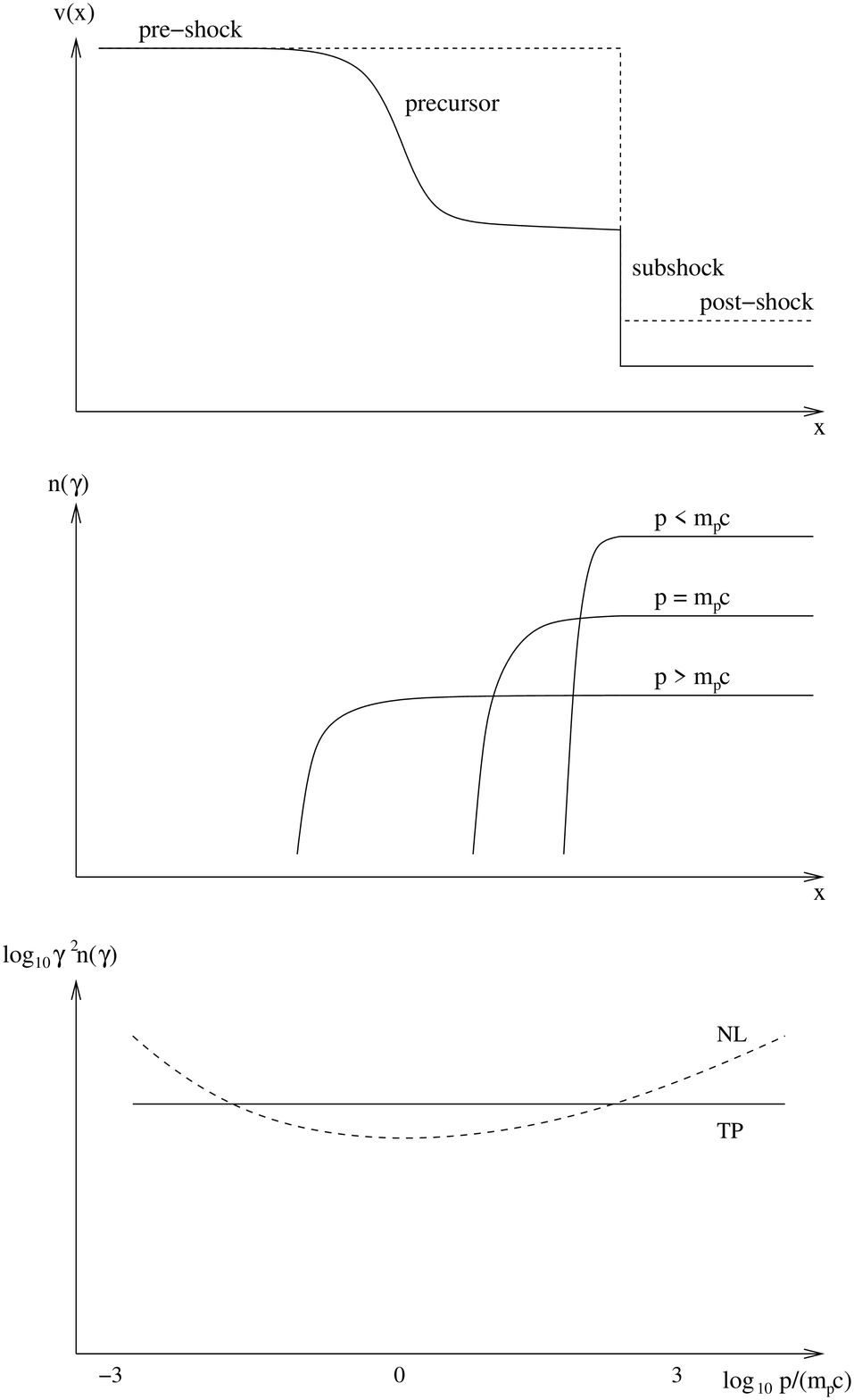,width=8.0cm}
\caption[]{Top: Schematic structure of a strong hydrodynamic shock. If
DSA places a negligible amount of energy into accelerated particles,
an ordinary discontinuous classical shock occurs (dashed line).  In
this case the undisturbed pre-shock flow is separated from the
post-shock flow by a discontinuity known as a subshock across which
the flow velocity decreases, and the density and pressure
increase. The scale length of the subshock, $l_{\rm s}$, is determined
by microphysical dissipation processes that result in gas heating. In
contrast, when particle acceleration is efficient, the diffusion of
non-thermal particles upstream of the subshock exerts a back-pressure
on the pre-shock flow, causing the gas velocity to decrease prior to
the subshock in a smooth precursor (solid line). The scale length of
the precursor, $l_{\rm d}$ ($>> l_{\rm s}$), is approximately the
diffusion length of the highest energy ions.  Middle: The spatial
distribution of the number density of non-thermal particles as a
function of energy in the modified shock. The highest energy particles
are able to diffuse further upstream from the subshock, and see the
greatest velocity difference between the upstream and downstream
flow. Non-thermal particles in the precursor are typically scattered
back across the subshock, and either undergo further acceleration, or
escape downstream. Bottom: The non-thermal particle energy spectrum
multiplied by $\gamma^{2}$ to emphasize the spectral curvature of the
non-linear (i.e. efficient particle acceleration) case. When DSA
places relatively little energy into accelerated particles the
standard test particle spectrum is obtained, as shown here for $p=2$
(solid). In contrast, the modified shock produced when DSA is
efficient leads to a concave curvature in the non-thermal particle
energy spectrum \citep[dashed - see][and references
therein]{Ellison:2004}.}
\label{fig:shock_mod}
\end{center}
\end{figure}

There are additional ways in which the spectral index can vary from
the canonical strong shock limit of $p=2$. For instance, when particle
acceleration is efficient, the diffusion of non-thermal particles
upstream of the subshock exerts a significant back-pressure on the
pre-shock flow, causing the gas velocity to decrease prior to the
subshock in a smooth precursor (Fig.~\ref{fig:shock_mod}). As the
highest energy particles are able to diffuse further upstream from the
subshock, and see the greatest velocity difference between the
upstream and downstream flow, a concave curvature to the non-thermal
particle energy spectrum is created. In addition, the velocity jump
across the subshock in this situation is much less than in the
classical shock, which results in a reduced post-shock thermal
temperature. Convincing evidence for such ``shock modification'' is
provided by analysis of observations of several young SNRs,
including Tycho \citep{Volk:2002} and SN~1006 \citep{Bamba:2003,
Berezhko:2003}, and from direct spacecraft measurements of the Earth's
bowshock and interplanetary shocks \citep[see references
in][]{Ellison:2005}.
Non-thermal ions are believed to be primarily responsible for
modifying shocks because they are thought to dominate the non-thermal
energy density (see Sec.~\ref{sec:norm}). Since the diffusion length
of electrons is much smaller than ions for a given Lorentz factor,
non-thermal electrons are confined to a region much closer to the
subshock. Thus, only those electrons with momenta $\gtsimm {\rm
m_{p}}c$ (i.e.  with Lorentz factors $\gtsimm 2000$) are able to
diffuse far enough upstream to see the velocity difference required to
obtain a spectral index $p<2$ \citep[assuming $p=2$ for the unmodified
shock - see Fig.~1 in][]{Ellison:2004}.  In CWB systems, the
synchrotron emission at GHz frequencies is from electrons with
relatively low momentum.  For example, at the apex of the WCR in
model~B of this work, electrons with $\gamma=50$ (i.e. momentum
$p=50\;m_{\rm e}c \approx 0.03\;{\rm m_{p}}c$) produce synchrotron
emission which peaks at $\nu=0.29\nu_{\rm c}\approx3.6$~GHz, where
$\nu_{\rm c}$ is the frequency where the synchrotron spectrum
cuts off \citep{Pacholczyk:1970}.  Hence, if shock modification occurs in
these systems, the synchrotron emission should be characterized by
$p>2$, though if re-acceleration occurs this is not
necessarily the case.


The spectral index of non-thermal particles can be modified if they
pass through a sequence of shocks. In the limit
of many shocks, the spectral index approaches $p=1$, independent of 
the shock strength. This process is known as ``re-acceleration''
\citep[see][and references therein]{Pope:1994}, and if the particles
suffer energy losses between the accelerating shocks, $p$ can become
even smaller \citep{Schlickeiser:1984,Melrose:1997}. Shock modification 
and particle re-acceleration are not necessarily mutually exclusive.
 
Some previous works \citep{Kirk:1989,Naito:1995} have also claimed
that $p<2$ can be obtained if the shock is oblique (i.e. when the B-field
is at some angle to the shock normal), but these rely on a
non-standard shock acceleration model, such as non-isotropic
scattering, and we do not consider this further.  Mechanisms
which increase the effective Mach number of the shock may also yield
$p<2$ \citep[e.g.,][]{Malkov:2000}, while $p>2$ may be obtained if
sub-diffusive cross-field transport is important
\citep{Duffy:1995,Kirk:1996}.

\subsubsection{Summary of assumed NT particle spectrum}
From energy considerations, DSA and reconnection appear to be the most
efficient mechanisms, with DSA likely having the edge \citep[see
also][]{Bednarek:2005}. Therefore, for the remainder of this work we
assume that the non-thermal particles (electrons and ions) are created
by DSA at the global shocks bounding the WCR and have power-law energy
distributions. We further assume that the downstream magnetic field 
is highly tangled. While we do not model the actual acceleration process
itself, this is consistent with the Bohm limit where the magnetic
fluctuations are of the same strength as the background field i.e.
$\delta B/B \sim 1$. The strong cross-field diffusion which results in this
limit allows efficient injection at quasi-perpendicular shocks.

The diffusion length scale of accelerating particles is given by
$l_{\rm diff} = \kappa/v_{\rm shk}$, where $v_{\rm shk}$ is the shock
speed, $\kappa = \lambda v/3$ is the spatial diffusion coefficient,
$\lambda \approx 3 r_{\rm g}$ is the mean-free-path, and $r_{\rm
g}=\gamma m c^{2}/qB$ is the gyroradius of particles with Lorentz
factor $\gamma$, mass $m$, and charge $q$ in a magnetic field $B$.
The acceleration of non-thermal electrons at the shocks is not
affected by the flow geometry of the downstream plasma, as the
diffusion lengthscale is only a small fraction of the width of the WCR
(e.g., $l_{\rm diff} \sim 10^{10}\;{\rm cm}$ for electrons with
$\gamma = 10^{5}$ at the shock apex of model~B, whereas the distance
between the global shocks and the contact discontinuity is $\approx 2
\times 10^{13}\;{\rm cm}$). Therefore, the geometry of the
post-shock flow will have no impact on our calculation of the
synchrotron and IC emission, or our finding that the electron
spectrum has $p < 2$. The flow geometry may become important
for non-thermal ions, but for protons and He nuclei this is only
likely to be the case for $\gamma \gtsimm 10^{4}$.  Detailed work
outside the scope of this paper is required to examine the effect of
the flow geometry on our predictions of the high energy emission from
the decay of neutral pions.

We do not consider the inverse-Compton (IC) cascade of
secondary electrons and positrons that occur when the optical depth of
$\gamma$-rays to pair-production in a stellar radiation field is high
and when IC scattering is efficient, though in WR\thinspace140 the
secondary $\gamma$-rays produced in the cascade and which subsequently
escape the system may be comparable to the observed flux of primary
$\gamma$-rays above GeV energies for the orbital phase considered in
this work \citep[cf.][]{Bednarek:2006}. The non-thermal electrons
formed by hadronic collisions, and the knock-on electrons formed when
non-thermal nuclei collide with thermal (atomic and free) electrons,
are also ignored in this work, though the emission resulting from
these processes is not expected to be dominant \citep[see
also][]{Mastichiadis:1996, Baring:1999,Berrington:2003}. We also
ignore an additional (though likely small) contribution to the
non-thermal radio emission from particles accelerated at wind-embedded
shocks caused by the line-deshadowing instability in radiatively
driven winds, since in WR\thinspace140 the non-thermal emission is
clearly related to binarity. We set $p$ as a free parameter for the
particle distribution immediately downstream of the shocks, with the
assumption that it is spatially invariant along the face of the
shocks, and identical for electrons and ions. For the ions $p$ does
not change in the post-shock flow, but as the non-thermal electrons
advect downstream IC cooling modifies their energy distribution, and
in this case $p$ is dependent on both their spatial position and
$\gamma$ \citep{Pittard:2006}. For instance, in model~B, electrons
accelerated at the WR shock on the line of centers between the stars
cool to $\gamma \ltsimm 100$ within a time of $\approx 6.3 \times
10^{4}\;{\rm s}$, during which they are advected a distance of
$\approx 4.7 \times 10^{12}\;{\rm cm}$ ($0.015 \;D$) downstream
(assuming that the speed of the post-shock flow is $0.25 \;v_{\infty}$
in this region of the WCR).  Electrons accelerated at a point on the
WR and O shocks which is a distance $D$ from the O star cool
to $\gamma \ltsimm 1000$ within a time of $\approx 2 \times
10^{5}\;{\rm s}$. During this time they are advected a distance of
$5.9 \times 10^{13}\;{\rm cm}$ ($0.19 \;D$) downstream (assuming a
post-shock flow speed close to the terminal wind speeds).

\subsection{Normalization}
\label{sec:norm}
\subsubsection{The non-thermal particle spectrum}
\label{sec:norm_part}
Since the population of non-thermal particles is not determined from
first principles in our models, the non-thermal electron energy
density, $U_{\rm rel,e}$, is assumed to be proportional to the thermal
particle internal energy density, $U_{\rm th}$ i.e. $U_{\rm rel,e} = 
\zeta_{\rm rel,e} U_{\rm th}$. Also, it is assumed that
shock obliquity does not influence the normalization of the
non-thermal electron energy density.  However, we note that in
general, the higher the obliquity, the lower the injection efficiency
\citep[although particles are actually accelerated more rapidly as a
result of shock drift along the surface and slower diffusion in the
shock normal direction - see,
e.g.,][]{Ellison:1995,Ellison:1996,Ellison:1999}.  In wide binaries,
such as WR\thinspace140, the shocks near the head of the WCR should,
in principle, be highly oblique. Yet this picture is further
complicated if the level of magnetic turbulence is at the Bohm limit,
since then the shock does not have a well-defined obliquity, and
strong cross-field diffusion results in efficient injection \citep[see
Fig.~6 in][]{Ellison:1995}. Such concerns are beyond the scope of the
present work.

To obtain an estimate of the $\gamma$-ray flux from the decay of
neutral pions, the energy density of relativistic ions must be
specified. Similar to the non-thermal electrons, it is assumed that
$U_{\rm rel,i} = \zeta_{\rm rel,i} U_{\rm th}$. Fits to the radio data
are only sensitive to $\zeta_{\rm rel,e}$ and constraining $\zeta_{\rm
rel,i}$ requires unambiguous detection of the emission from
neutral pion-decay, which has not been attained for CWB
systems. Moreover, the relative injection efficiency of electrons and
ions is not known, while the value of $\zeta_{\rm rel,i}/\zeta_{\rm
rel,e}$ also depends on the spectral index $p$
\citep{Schlickeiser:2002}.  Having said this, it is reasonably certain
that DSA transfers considerably less energy into non-thermal electrons
than into non-thermal ions \citep[\eg,][]{Baring:1999}, while the
relative difficulty of accelerating electrons is also consistent with
the well established value for the ratio of proton to electron energy
densities for Galactic cosmic rays of $\sim 100$
\citep{Longair:1994}. In this work we somewhat arbitrarily assume that
$\zeta_{\rm rel,i} = 100\zeta_{\rm rel,e}$, unless otherwise noted,
though the actual ratio in WR\thinspace140 could differ
significantly. As a result, our predictions of the $\gamma$-ray flux
from neutral pion decay and neutrino flux from charged pion decay are
highly uncertain. However, hard upper limits for these fluxes can be
obtained due to the requirement that the total power put into
non-thermal particles must be less than the available kinetic power in
the WCR i.e. $\zeta_{\rm rel,e} + \zeta_{\rm rel,i} \ltsimm 1$.

As non-thermal particles flow downstream, they are subject to
adiabatic losses. This is achieved in an approximate fashion in this
work through the assumption that $U_{\rm rel,e}$ and $U_{\rm rel,i}$
are proportional to $U_{\rm th}$. In reality the adiabatic indices for
thermal and relativistic plasma differ, and adiabatic expansion also
reduces the Lorentz factor of the particles. While neither of these
effects is treated properly in our model, the specific
parameterization we have chosen will not lead to significant
errors in the resulting non-thermal radio emission. This is because
IC cooling is so rapid for high energy electrons that synchrotron
emission above $\sim1$~GHz is largely confined to a thin region near
the shocks, where the effects of adiabatic expansion of the flow are
minimal. Likewise, the radiation from pion decay is concentrated near
the apex of the WCR.

\subsubsection{The magnetic field}
\label{sec:norm_b}
If the field is frozen into the post-shock plasma of density $\rho$,
one would anticipate $B \propto \rho$ ($U_{\rm B} \propto \rho^{2}$),
but amplification of the magnetic field by the accelerated particles
\citep[e.g.,][]{Bell:2004} and/or reconnection complicates this simple
scaling. This also implies that it is not necessarily straightforward
to determine the B-field in the WCR from assumptions about the
strength of the surface magnetic field at each star. Indirect evidence
for amplification has recently been obtained in several SNRs
\citep[e.g.,][]{Berezhko:2003,Berezhko:2004,Ellison:2005}. Instead, for
convenience, we have assumed that the magnetic energy density is
proportional to the thermal particle energy density i.e.  $U_{\rm B} =
\zeta_{\rm B} U_{\rm th}$. This choice is equivalent to some
amplification in the downstream flow, but we have verified that there
is little difference in the shape of the synchrotron spectrum between
these two methods.  Compared to other work where the B-field is
assumed to be spatially invariant, this approximation at least has the benefit
that the B-field declines away from the symmetry axis.
Undoubtedly, a better description for the B-field will be obtained in
future MHD models of the WCR. Our final assumption is that
the post-shock B-field is highly tangled so that the synchrotron
emission is isotropic.

\subsubsection{Limits on $\zeta_{\rm B}$ and $\zeta_{\rm rel,e}$}
When modelling the radio emission from specific systems, $\zeta_{\rm
B}$ and $\zeta_{\rm rel,e}$ are chosen to best match the observed
radio emission. As $\zeta_{\rm B}$ decreases, the value of $\zeta_{\rm
rel,e}$ has to increase to reproduce an observed radio flux. A
lower limit on $\zeta_{\rm B}$ can be obtained from the fact that the
Razin turnover frequency, $\nu_{\rm R}$, increases with declining
$\zeta_{\rm B}$ ($\nu_{\rm R} \approx 20n_{\rm e}/B$, where $n_{\rm
e}$ is the electron number density and the magnetic field,
$B=\sqrt{8\pi U_{\rm B}}$), or because the value of $\zeta_{\rm
rel,e}$ required to match an observed flux exceeds the available
power for particle acceleration in the WCR.  
An upper limit on $\zeta_{\rm B}$ is obtained from the
necessary conditions for particle acceleration
\citep[see][]{Eichler:1993}.  Since there is currently little guidance
for the ratio of $\zeta_{\rm B}/\zeta_{\rm rel,e}$ in CWB systems, we
consider this to be a free parameter in our models, subject to the
above constraints.

\subsection{Maximum energy of non-thermal particles}
The maximum value of $\gamma$ for electrons is obtained by balancing
the rate of energy increase from DSA to the rate of energy loss by IC
cooling \citep[e.g., see Eqs.~4 and 7 in][]{Pittard:2006}. This is
spatially dependent due to the geometric dilution of both the stellar
radiation and magnetic fields, and also dependent on orbital phase for
binaries with eccentric orbits like WR\thinspace140. An evaluation of
these equations at the points where the shocks intersect the line of
centers through the stars reveals that $\gamma_{\rm max} \sim 10^{5}$
at orbital phase 0.837 for sensible values of $B$. Most of the wind
power passing into the WCR occurs within an off-axis distance of $\sim
2 r_{\rm O}$ from the line of centers \citep{Pittard:2002b}, and one
expects that most of the energy transfer into non-thermal particles
must occur within this volume too. The value of $\gamma_{\rm max}$
evaluated at a position on the shock front at the extremity of this
region is only 25 per cent higher than on the line of centers if the
magnetic field strength is assumed to decline as $1/r$ and if there is
no dependence on the shock obliquity (as already mentioned, this is
not well defined if the level of magnetic turbulence is at
the Bohm limit).  Therefore, for simplicity, throughout this work it
is assumed that $\gamma_{\rm max}{\rm (electrons)}=10^{5}$ along the
entire shock fronts.

While the above evaluation was for the Thomson limit, at such energies
IC losses occur in the Klein-Nishina regime, and the spectral
distribution as well as the cut-off energy will change. However, in
the calculations in this paper, IC cooling is so rapid near the
apex of the WCR that the downstream population of 
non-thermal electrons in this region is essentially independent of 
whether Thomson or Klein-Nishina cross-sections are used. Thus, the IC
emission and our predictions for the $\gamma$-ray fluxes are essentially 
identical too, with the exception that the high energy cutoff is  
sensitive (by a factor of order 2) to such details.


The maximum non-thermal ion energy is set by balancing the
timescale of their acceleration to their advection timescale out of
the system.  Following \citet{Bednarek:2005}\footnote{In Eq.~3 of
\citet{Bednarek:2005} the denominator should have a factor $c^{3}$ as
shown here in Eq.~\ref{eq:gam_max_ions}.},
\begin{equation}
\label{eq:gam_max_ions}
\gamma_{\rm max}{\rm (ions)} = \frac{3ZeBr_{\rm O}v}{c^{3}A{\rm m_{p}}}
\end{equation}
where $Ze$ and $A{\rm m_{p}}$ are the charge and mass of the ion, and
$B$ and $v$ are the pre-shock magnetic field and wind speed.  For wide
binaries where the pre-shock magnetic field declines as $1/r$,
$\gamma_{\rm max}{\rm (ions)}$ is expected to be independent of $D$.

For the models considered in this paper, $\gamma_{\rm max}{\rm (ions)}
\sim 10^{5}$, so for convenience, we set $\gamma_{\rm max} = 10^{5}$
for both the electron and ion distributions in the remainder of this
work.
The predicted $\gamma$-ray flux is, of course, sensitive to the value
of $\gamma_{\rm max}$ chosen for both the electrons and ions.
In addition, $\gamma_{\rm min}=1$ is assumed.

\subsection{Composition of non-thermal nuclei}
In the simplest test-particle calculations of DSA i.e. with no feedback, 
all particles accelerated in a given shock will have identical
power-law spectral shapes in momentum, regardless of their charge, and
the non-thermal particle composition reflects the composition of
thermal particles. However, in non-linear models, the spectral
shape and injection and acceleration efficiencies are dependent on the
ionic species.  Ions with large mass-to-charge (A/Q) ratios diffuse
further upstream than protons of the same energy per nucleon (provided
that both are non-relativistic), and are accelerated more efficiently
and obtain a flatter spectrum because they ``see'' a larger velocity
difference and are more easily injected to suprathermal energies.
Among the volatile elements, observational support comes from 
Galactic cosmic rays, which show an enhancement
of heavier elements relative to lighter ones \citep{Meyer:1997}.
A general enhancement of the refractory elements relative to the
volatile ones is also apparent from the Galactic cosmic ray composition
\citep{Meyer:1997}. This has been explained by acceleration
of dust grains (which can have huge mass-to-charge ratios of 
$\sim 10^{4}-10^{8}$), followed by the sputtering of atoms from these grains.
However, the harsh UV radiation fields in CWB systems makes it
unlikely that dust grains can form near the head of the WCR, 
so an enhancement of refactory elements is not expected.
Several CWB systems, including WR\thinspace140, show observational 
signatures of either episodic or continuous dust formation 
\citep[e.g.,][]{Williams:1996,
Tuthill:1999,Marchenko:2002}, though in all cases the
dust is thought to form well downstream of the apex of the WCR.
%
%
Observations have further revealed that the relative abundance of
heavy ions increases with energy \citep[see][]
{Bykov:2001,Hillas:2005}. This is supported by the results from
non-linear models where variation with shock speed and Mach number is
also seen \citep{Ellison:1997}.  It is also possible that
shock obliquity may influence the non-thermal particle composition.

High energy heavy nuclei may also lose nucleons due to collisions with
stellar photons i.e. by photo-disintegration.  This occurs if the
photon energy in the reference frame of the nuclei, $E_{\gamma} = 3
k_{\rm B}T \gamma (1 + {\rm cos}\;\theta)$, exceeds $\sim 2$~MeV
\citep{Bednarek:2005}.  Here $T$ is the temperature of the radiation
field. In WR\thinspace140, this process applies to nuclei with
$\gamma \gtsimm 8.5 \times 10^{4}$, and is only likely to
affect a small part of the top end of the non-thermal particle
distribution. In addition, an estimate of the ratio of the
characteristic photo-disintegration timescale to the acceleration and
advection timescales reveals that photo-disintegration is not
efficient for particles with $\gamma\ltsimm 10^{5}$ in WR\thinspace140
at the orbital phase considered in this work. The fragmentation
of high energy nuclei through collisions with thermal ions is likewise
negligible.

Given the uncertainties discussed earlier, we adopt the following
simplifications: i) identical spectra for all non-thermal particles,
ii) identical composition to that of the thermal particles, iii) no
dependence on the shock obliquity.

\section{Non-thermal X-ray and $\gamma$-ray Emission and absorption}
\label{sec:gamma}
The presence of non-thermal electrons and ions should give rise
to X-ray and $\gamma$-ray emission from several separate mechanisms,
including IC scattering, relativistic bremsstrahlung, and pion decay
(the non-thermal electrons are not energetic enough to produce
synchrotron X-ray emission).  While definitive evidence for
non-thermal X-ray and $\gamma$-ray emission from CWBs does not yet
exist, many of the yet unidentified EGRET sources appear correlated
with populations of massive stars \citep{Romero:1999}, and some
systems appear to show a power-law tail extending to higher energies
than the thermal X-ray emission \citep{Viotti:2004,DeBecker:2004b}.  A
critical question is whether the non-thermal X-ray and $\gamma$-ray
fluxes from CWB systems are sufficiently high to be detected with
instruments onboard current observatories,
and instruments on observatories that will be
operational in the near future.


Previous predictions of the IC emission have usually been based on
the fact that the ratio of the luminosity from IC scattering to
the synchrotron luminosity is equal to the ratio of the photon energy
density, $U_{\rm ph}$, to the magnetic field energy density, $U_{\rm
B}$:
\begin{equation}
\label{eq:icsynclum}
\frac{L_{\rm ic}}{L_{\rm sync}} = \frac{U_{\rm ph}}{U_{\rm B}}. 
\end{equation}
Typically, the ratio of
$U_{\rm ph}/U_{\rm B}$ is evaluated only at the stagnation point,
whereas in reality the emission is generated throughout the
WCR, with $U_{\rm ph}/U_{\rm B}$ varying spatially. In
addition, $L_{\rm sync}$ is usually set to the {\em observed}
synchrotron luminosity, whereas free-free absorption by the extended
wind envelopes can be significant \citep[see,
\eg,][]{Pittard:2006}. In such cases the intrinsic synchroton
luminosity, and consequently the non-thermal X-ray and $\gamma$-ray
luminosity, will be underestimated if radio data, particularly at
lower frequencies, are used without regard to free-free absorption
(cf. Figs.~\ref{fig:var_theta}, \ref{fig:ffabs_components} and 
\ref{fig:razin_components}). Furthermore, the predictive power of
Eq.~\ref{eq:icsynclum} is greatly undermined by the fact that the
magnetic field strength in the WCR is generally not known with any
certainty. Since $U_{\rm B} \propto B^{2}$, small changes in the
estimated value of the magnetic field stength, $B$, can lead to large 
changes in $L_{\rm ic}$ \citep[see Fig.~3 in][]{Benaglia:2003}.

To obtain a more robust estimate of the IC emission from CWB systems,
the population and spatial distribution of non-thermal electrons must
be determined directly. This information can, in principle, be
acquired from model fits to radio data, since the IC emission arises
from the same non-thermal electrons responsible for the synchrotron
radio emission. As already noted, an improved theoretical basis for
the modelling of the radio emission from such systems was developed by
\citet{Dougherty:2003} and \citet{Pittard:2006}, and we use this as
the foundation for our calculations of the non-thermal X-ray and
$\gamma$-ray emission.  Key benefits over previous work include a more
realistic description of the geometry of the WCR, the evolution of the
energy distribution of the non-thermal electrons as they advect
downstream from their acceleration site, and absorption of high energy
photons by pair production in the stellar radiation fields.  Our
predictions for the expected $\gamma$-ray emission are based on model
fits to the available radio and X-ray data, which allow us to
constrain the range of free parameters and build in a degree of
consistency for our estimates.

\subsection{IC emission}
\label{sec:ic}
IC cooling is a major energy loss mechanism for non-thermal electrons
in CWB systems, and the non-thermal X-ray and $\gamma$-ray luminosity 
should exceed the synchrotron radio
luminosity by several orders of magnitude. In the Thomson regime, the
average energy of a photon after isotropic IC scattering is 
\citep{Blumenthal:1970}
\begin{equation}
h\nu_{\rm IC} \sim \frac{4}{3} \gamma^{2} h\nu_{*}.
\end{equation} 
Since the average photon energy of an early-type star, $h\nu_{*} \sim
10\;$eV, Lorentz factors of order $10-10^{4}$ are sufficient to
produce IC X-ray and $\gamma$-ray radiation up to GeV energies. The
resulting emission has a spectral shape which is identical to the
intrinsic synchrotron emission at radio frequencies.  For our
calculations, we assume that the scattering is isotropic and takes
place in the Thomson limit. While a more thorough treatment would take
account of Klein-Nishina effects, which are important at high
energies, recent work has shown that the resulting change to the
non-thermal electron energy distribution is rather small when the
stellar separation is $\gtsimm 2 \times 10^{14}\cm$
\citep{reimer:2006}. This condition is satisfied in WR\thinspace140
between orbital phases $\phi=0.1-0.9$, and also in wider binaries like
WR\thinspace146 and WR\thinspace147.

Consideration should also be given to the anisotropic nature of the IC
process, where the emitted power is dependent on the scattering
angle. However, if the O star is in front of the WR star (which is the
case for WR\thinspace140 at $\phi=0.837$ - see Sec.~\ref{sec:wr140}),
the variation in the resulting emission is less than a factor of 2.5
\citep{reimer:2006}. In Sec.~\ref{sec:calc_em}, we show that
the other uncertainties affecting the IC emission, such as the value
of $\zeta_{\rm e}$ determined from fitting the radio data, are at
least comparable in magnitude, so we leave this issue to future work.

\subsection{Relativistic Bremsstrahlung}
\label{sec:brem}
Bremsstrahlung radiation is produced when a charged particle is
accelerated in the Coulomb field of another charged particle. It is
possible to obtain emission at $\gamma$-ray energies if one of the
particles is relativistic, since photons of comparable energy to that
of the emitting particle can be produced. Inverse bremsstrahlung,
produced by collisions between high energy ions and thermal electrons,
is an insignificant contribution to the total emission unless the
ratio of the energy density of non-thermal ions to electrons is very
high. Therefore, we concentrate on emission produced by collisions
between non-thermal electrons and thermal electrons and ions.  A more
detailed discussion can be found in \cite{Baring:1999}.
 
The rate of photon production in the energy interval between
$\epsilon_{\gamma}$ and $\epsilon_{\gamma} + d \epsilon_{\gamma}$ by an
electron of kinetic energy $E_{\rm e}$ is
\begin{equation}
\label{eq:ep}
\frac{dn_{\gamma}(E_{\rm e},\epsilon_{\gamma})}{dt} = 
v_{\rm e}\left[n_{\rm e}\sigma_{\rm ee}(E_{\rm e},\epsilon_{\gamma}) +
\sum_{i} n_{i}Z_{i}^{2}\sigma_{\rm ep} 
(E_{\rm e},\epsilon_{\gamma})\right],
\end{equation}
where $\epsilon_{\gamma}$ is the photon energy in units of $m_{\rm
e}c^{2}$, $v_{\rm e}$ is the relative velocity of the colliding
particles, $n_{i}$ and $n_{\rm e}$ are the number density of ions
and electrons respectively, and $Z_{i}$ is the charge of the ion. The
electron-proton cross section, $\sigma_{\rm ep}$, is the
Bethe-Heitler cross section evaluated in the Born approximation. 
We determine $\sigma_{\rm ep}$ for any electron energy,
relativistic or non-relativistic, using equation 3BN in \cite{Koch:1959},
and omit the Coulomb corrections at low energies. The
cross section for electron collisions with fully ionized ions has a 
$Z^{2}$ charge dependence, accounting for the factor in Eq.~\ref{eq:ep}.
For the electron-electron cross section, $\sigma_{\rm ee}$, the 
approach noted in \cite{Baring:1999} is followed. 

For $\gamma_{\rm e} \gg 1$ and $\epsilon_{\gamma} \gg 1$, the cross
sections depend only on the particle's charge and not its mass, so
$\sigma_{\rm ep} = \sigma_{\rm ee}$. The ratio of the
electron-electron to electron-ion contributions is then simply 
\begin{equation}
\label{eq:pp}
\frac{n_{\rm e}}{\sum_{i}Z^{2}_{i}n_{i}}. 
\end{equation}
For material with solar abundances this ratio is $\approx 0.86$, while
in the shocked WR wind of WR\thinspace140 it is $\approx 0.47$ (see
Sec.~\ref{sec:wr140}). At high energies, the bremsstrahlung spectrum
(in units of $\phpscm2$) has a spectral index which reflects that of
the non-thermal particles i.e. $p$.

\subsection{$\pi^{0}$ decay}
\label{sec:pi}
High energy hadronic collisions can create neutral pions which
subsequently decay into two $\gamma$-ray photons, \eg, $p + p
\rightarrow \pi^{0} + X$, $\pi^{0} \rightarrow \gamma + \gamma$.  The
final stage has a branching ratio of $\approx 98.8\%$ (the decay
$\pi^{0} \rightarrow \gamma + e^{+} + e^{-}$ accounts for the other
1.2\%).  The pion decay process yields information on the population
of non-thermal nucleons, in contrast to the IC and bremsstrahlung
processes where the emission is dependent on the population of
non-thermal electrons. 

The cross-section for hadronic collisions
depends on the ionic species involved, so we first consider the
$\pi^{0}$ emissivity from proton-proton collisions. The production
rate for neutral pions of energy $E_{\pi}$ per thermal proton can be
easily evaluated using a $\delta$-functional approximation for the
differential cross-section \citep{Aharonian:1996}:
\begin{equation}
\label{eq:pion1}
q_{\pi}(E_{\pi}) = \int \delta(E_{\pi} - f_{\pi}E_{\rm k}) 
\sigma_{\rm pp2}(E) J(E) dE,
\end{equation}
where $E_{\rm k}$ is the kinetic energy of the non-thermal protons,
$f_{\pi}$ is the fraction of the kinetic energy of the non-thermal
proton which is transferred to the $\pi^{0}$ meson, $\sigma_{\rm pp2}$
is the total inelastic cross section \citep{Aharonian:1996}, and
$J(E)$ is the flux of non-thermal protons of total energy $E$ (${\rm
protons}\s^{-1}\cm^{-2}\GeV^{-1}$). Here, only the contribution to the
$\gamma$-ray emission from the neutral pion is considered. After the
interaction the non-thermal proton keeps approximately half of its
energy, while the other half goes to pions. On average the neutral pion gets
approximately one third of this energy, or approximately
one sixth of the energy of the non-thermal proton, so
$f_{\pi}=0.17$. 
Thus, Eq.~\ref{eq:pion1} becomes (${\rm pions}\s^{-1}\GeV^{-1}{\rm H^{-1}}$):
\begin{equation}
\label{eq:pion2}
q_{\pi}(E_{\pi}) = \frac{1}{f_{\pi}} \sigma_{\rm pp2}(E) J(E).
\end{equation}
The $\gamma$-ray emissivity (${\rm photons}\s^{-1}\GeV^{-1}{\rm H^{-1}}$) 
is then
\begin{equation}
\label{eq:pion3}
q_{\gamma}(E_{\gamma}) = 2 \int^{\infty}_{E_{\rm min}} 
\frac{q_{\pi}(E_{\pi})}{\sqrt{E_{\pi}^{2} - {\rm m_{\pi}}^{2} c^{4}}} dE_{\pi},
\end{equation}
where $E_{\rm min}=E_{\gamma}+{\rm m_{\pi}}^{2}c^{4}/4E_{\gamma}$ and ${\rm m_{\pi}}$
is the mass of the neutal pion (134.9766 MeV/${\rm c^{2}}$).

In CWB systems, we must account for the composition of non-thermal and
thermal ions. For collisions involving two nuclei with atomic weights
$A_{1}$ and $A_{2}$, the cross section is \citep{Orth:1976}
\begin{equation}
\label{eq:pion4}
\sigma \sim (A_{1}^{3/8} + A_{2}^{3/8} - 1)^{2} \sigma_{\rm pp2}.
\end{equation} 
This prescription is appropriate for proton-helium interactions, but
its validity for heavy nuclei, such as Fe, is unclear. It is adopted
here since its potential inaccuracy is small compared to the other
uncertainties and assumptions in this work. In the case of WR stars,
such as WR\thinspace140, the wind contains little or no hydrogen, and
the cross-section for pion decay is best calculated relative to the
He-He cross section $\sigma_{\alpha\alpha}=5.6\sigma_{\rm pp2}$.

Considering only protons and Helium
nuclei, an enhancement to the pion production rate in
Eq.~\ref{eq:pion2} of 1.61 is obtained for material of solar
composition.  Greater enhancements are obtained for the nuclear
processed material in WR winds. For example, if collisions involving
He, C, O and Ne are considered for the WC7 wind abundances in
WR\thinspace140 (see Sec.~\ref{sec:composition}), then $\sigma \sim
1.4 \sigma_{\alpha \alpha} \sim 7.8 \sigma_{\rm pp2}$.

\subsection{$\gamma$-ray absorption}
\label{sec:gamma_abs}
$\gamma$-rays above a suitable threshold energy may create
electron-positron pairs via interaction with a lower energy photon or
with charged nuclei. While the latter process has an insignificant
optical depth in CWB systems, in contrast the high stellar radiation
energy density may be a significant source of opacity to high energy
$\gamma$-rays. The probability of absorption depends on the cosine of
the angle between the directions of the two photons, $\mu$.  For a
high energy photon of energy $E_{\gamma}$ interacting with a stellar
photon of energy $\epsilon$, the optical depth is given by
\begin{equation}
\label{eq:gamma_tau}
\frac{d\tau}{dr}=\int\sigma(\chi)(1-\mu)n(\epsilon,\mu,r) d\epsilon d\mu,
\end{equation}
where $\chi = \sqrt{E_{\gamma}\epsilon(1-\mu)/2}$ is the centre-of-momentum
frame energy\footnote{The equation for $\chi$ contains a typo in 
\cite{Baring:1997}.} scaled by $m_{\rm e}c^{2}$, 
\begin{eqnarray}
& \sigma(\chi)= & \frac{\pi r_{\rm e}^{2}}{\chi^{6}}\left[(2\chi^{4} +2\chi^{2}-1) \;{\rm ln}(\chi+\sqrt{\chi^{2}-1})\right. \nonumber \\
& & \left. - \chi(1+\chi^{2})\sqrt{\chi^{2}-1}\right],
\end{eqnarray}
where $r_{\rm e}=e^{2}/m_{\rm e}c^{2}$, and $n(\epsilon,\mu,r)$ is the 
number density of stellar photons. The threshold for pair production, 
and thus for absorption, is $\chi \geq 1$.

\begin{figure}
\begin{center}
\psfig{figure=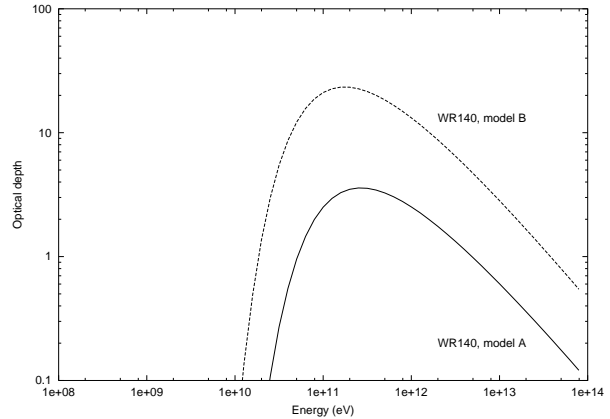,width=8.0cm}
\caption[]{The optical depth from two-photon pair production as a
function of the non-thermal photon energy for a single line-of-sight
from the stagnation point of the WCR in WR\thinspace140. The shape of
the curves reflects the Planck distribution for the stellar photons
($T = 45,000\;{\rm K}$). We show calculations at $\phi=0.837$ for
models~A and B (see Sec.~\ref{sec:wr140}). In each case, only the
photons from the O star are considered.  The energy of peak absorption
is a function of the viewing angle, $\theta$, and increases as
$\theta$ decreases.}
\label{fig:gamma_tau}
\end{center}
\end{figure}

In wide CWBs it can be assumed that the stars are point-like, and that
the stellar photons stream radially away from each star. This allows
the $(1-\mu)$ term to be taken out of the integral in
Eq.~\ref{eq:gamma_tau}. The optical depth as a function of non-thermal
photon energy from the stagnation point to an observer for two models
of WR\thinspace140 is shown in
Fig.~\ref{fig:gamma_tau}. Table~\ref{tab:wr14067_params} summarizes
some key parameters of these models.

Of the two models shown in Fig.~\ref{fig:gamma_tau}, both the shock
apex and the line of sight from the apex are closer to the O star in
model~B. The greatest opacity occurs at $\sim 100$~GeV, where optical
depths as high as $\sim 25$ (model~B) may be reached.  At lower and
higher energies the optical depth decreases, and optical depth unity
occurs at approximately 50~GeV and 5~TeV for model~A, and at
approximately 20~GeV and 40~TeV for model~B. In practice, the
effective optical depth is likely to be somewhat smaller due to the
spatial extension of the emission region. More accurate calculations
using radiative transfer along multiple sight lines are presented in
Sec.~\ref{sec:high_en_em}.  The CWB system WR\thinspace146, which has
a wider stellar separation, is expected to have a maximum optical
depth to pair-production of $\sim 1$. In WR\thinspace147, which is
wider still, the attenuation is negligible.  Similar conclusions have
also been noted by \cite{reimer:2006}.

\section{Non-thermal emission models of WR\thinspace140}
\label{sec:wr140}
In this section we apply our model to WR\thinspace140
(HD\thinspace193793), the archetype of long-period CWB
systems. WR\thinspace140 consists of a WC7 star and an O4-5 star in a
highly elliptical orbit ($e \approx 0.88$), and is well known for the
dramatic variations in its emission from near-IR to radio wavelengths
\citep{Williams:1990,White:1995}, and also at X-ray energies during
its 7.9-year orbit \citep{Zhekov:2000,Pollock:2002,Pollock:2005}.  The
variability appears to be linked to the WCR, which experiences
significant changes as the stellar separation varies between $\sim
2$~AU at periastron and $\sim 30$~AU at apastron.  The observed radio
emission increases by up to two orders of magnitude between periastron
and a frequency-dependent peak between orbital phases 0.65 to 0.85,
followed by a steep decline.  The decline of the radio emission prior
to periastron is consistent with the extinction suffered by the X-ray
emission at these orbital phases.
The X-ray lightcurve does not follow the expected $1/D$ variation
for an adiabatic WCR \citep{Pollock:2002}, which may indicate that the
WCR becomes radiative near periastron \citep{Varricatt:2004}. It is
also possible that this discrepancy is caused by a substantial amount
of the available energy in the wind-wind collision being placed into
non-thermal particles, with the efficiency varying as a function of
phase, though we believe this scenario should be considered as less likely.

WR\thinspace140 is also noteable for its possible association with an
unidentified EGRET source, lying on the outskirts of the positional
error box of 3EG~J2022+4317 \citep{Romero:1999}. The error box has a
radius of $\approx 0.7^{\circ}$, but WR\thinspace140 is the only known
high-energy source in the vicinity. 3EG~J2022+4317 has a flux of
$(24.7 \pm 5.2) \times 10^{-8}\;\phpscm2$, and a photon spectral index
$\Gamma=2.31\pm0.19$ (where $N(E) \propto E^{-\Gamma}$)
\citep{Hartman:1999}.  The variability of 3EG~J2022+4317 is discussed
in some detail by \citet{Benaglia:2003}. Since it was only detected in
3 out of 12 observing periods (at approximate orbital phases 0.97, 1.00, and
1.16), the upper limits of the remaining observations may hide real
variability.

It has been suggested previously that better
agreement may be obtained between observed and modelled radio lightcurves 
if the wind of the WR star is disc-like
\citep{White:1995}, but a specific (and unusual) orientation is
required for consistency with polarization measurements
\citep[see][and discussion therein]{Marchenko:2003}, and there is no 
evidence for such a disk in
X-ray observations \citep{Pollock:2002}. In the following, it is
assumed that the undisturbed wind is isotropic.

\subsection{Details of the model}
Our calculations of the observed thermal and non-thermal emission from
CWB systems are based on the density and temperature distribution
obtained from 2D hydrodynamical simulations of the stellar winds and
their collision \citep[see][]{Dougherty:2003,Pittard:2006}. At radio
frequencies, both the stellar winds and the WCR contribute thermal
emission and absorption. The free-free emission and absorption
coefficients are determined from the local temperature and density
values on the hydrodynamic grid, and clumping is accounted for (see
Sec.~\ref{sec:clump}). The non-thermal emission is
modelled after specifying the population and spatial distribution 
of non-thermal particles, as noted in Sec.~\ref{sec:ntpart_spec}.
Additional details relating to the hydrodynamical code, and the method
and assumptions used to obtain the various emission and absorption
coefficients in each cell, can be found in \citet[][]{Pittard:2006}
and references therein.

The assumption of axisymmetry in our 2D models is good until orbital
phases of $\sim 0.97$, as the orbital eccentricity is so high. Since
radiative braking \citep{Gayley:1997} or inhibition
\citep{Stevens:1994} is not expected to be important, the wind speeds
are assumed to be constant.  The calculations are performed on a grid
with $1600 \times 800$ cells, which spans the range $0 \leq r \leq 4
\times 10^{15} \cm$, $-4 \times 10^{15} \leq z \leq 4 \times 10^{15}
\cm$, and is large enough to capture the vast majority of the radio
and X-ray emission from the WCR,
though inevitably the models underestimate the radio flux at the
lowest frequencies considered. The line of sight into the system
is specified by the angle $\theta$, which is measured from the line
perpendicular to the line of centers of the two stars: $\theta =
+90^{\circ}$ corresponds to the O star in front of the WR star,
$\theta = 0^{\circ}$ corresponds to quadrature, and $\theta =
-90^{\circ}$ corresponds to the WR star in front of the O star.

\subsection{Parameters of WR\thinspace140}
\label{sec:wr140_parms}
Recent high-resolution VLBA observations at 8.4~GHz between orbital
phase 0.7 and 0.9 have provided important new constraints to models of
WR\thinspace140 \citep{Dougherty:2005}. An arc of emission is
observed, resembling the bow-shaped morphology expected for the WCR,
and rotates as the orbit progresses. This rotation allows derivation
of the orbit inclination and semi-major axis, leading to a geometric
distance determination ($1.85\pm0.16$~kpc) that is independent of
stellar parameters. This distance is greater than previously thought,
and implies the O star is a supergiant. Therefore, we adopt ${\rm
log}\;({\rm L_{bol}/\Lsol}) = 6.18$ and $v_{\infty} = 3100 \kmps$ for
the O4-5I primary, and ${\rm log}\;({\rm L_{bol}/\Lsol}) = 5.5$ and
$v_{\infty} = 2860 \kmps$ for the WC7 secondary
\citep{Dougherty:2005}.  These luminosities are higher than average
for the spectral types. \citet{Herrero:2002} and \citet{Repolust:2004}
suggest a typical value of 5.7-5.9 for an O4-5 supergiant, while the
mean luminosity for WC7 stars is about 0.4 dex lower (although the
scatter in the absolute magnitude of WC7's is $\sim1$
magnitude). Therefore, there may be some scope for lowering the
adopted luminosities.

\citet{Dougherty:2005} used the models of \citet{Zhekov:2000} of the
observed thermal X-ray luminosity to derive a mass-loss rate for the
WR star of $4.3 \times10^{-5} \;\Msolpyr$. However, there are several
complicating factors, such as the inadequate constraints on the
composition of the WC7 wind and on the wind momentum ratio, $\eta$. As
$\eta$ increases, the opening angle of the WCR increases, and a
greater fraction of the WC7 wind is shocked, resulting in an increase
in the X-ray luminosity if the mass-loss rate of the WR star is
unchanged.  There is also the possibility of some degree of shock
modification.  These issues are explored in the following
sub-sections, leading to improved estimates of the mass-loss rates and
$\eta$ in WR\thinspace140.

\begin{table}
\begin{center}
\caption[]{Some key parameters for our models of WR\thinspace140 at
$\phi=0.837$ (Sec.~\ref{sec:wr140}).  $D$ is the binary
separation, $\eta$ is the wind momentum ratio, $r_{\rm O}$ is the distance 
of the stagnation point of the WCR from the O star in units of the 
stellar separation $D$, and $d$ is
the distance of the system.  $U_{\rm th}$, $U_{\rm B}$, $U_{\rm ph}$,
$n_{\rm e}$, and $B$ are all evaluated at the stagnation point, while
the pre-shock Alfv\'{e}nic Mach number, $M_{\rm A}$, is evaluated 
immediately upstream of the WR shock on the
line of centers, assuming that at this point the shock is highly
perpendicular and that the B-field is reduced by a factor of 4 from its
post-shock value. $\theta$ is the angle of the line of sight into the system.
Two values are given for $n_{\rm e}$ - these are for the
WR and O star side of the contact discontinuity respectively.}
\label{tab:wr14067_params}
\begin{tabular}{lll}
\hline
\hline
Parameter & Model~A & Model~B\\
\hline
$D$ ($10^{14} \cm$) & \multicolumn{2}{c}{3.13} \\
${\rm log} \;(L_{\rm O,bol/\Lsol})$ & \multicolumn{2}{c}{6.18}\\
${\rm log} \;(L_{\rm WR,bol/\Lsol})$ & \multicolumn{2}{c}{5.5}\\
$d \;({\rm kpc})$ & \multicolumn{2}{c}{1.85} \\
$\eta$ & 0.22 & 0.02 \\
$r_{\rm O}$ & 0.32 & 0.124 \\
$\zeta_{\rm rel,e}$ & $1.38 \times 10^{-3}$ & $5.36 \times 10^{-3}$\\
$\zeta_{\rm B}$ & $0.05$ & $0.05$\\
$U_{\rm th} \;(\ergpcm3)$ & 0.78 & 1.2 \\
$U_{\rm B} \;(\ergpcm3)$ & 0.04 & 0.06 \\
$U_{\rm ph} \;(\ergpcm3)$ & 1.6 & 10.3 \\
$n_{\rm e}{\rm \;(WR)} \;(\pcm3)$ & $3.7 \times 10^{6}$ & $6.0 \times 10^{6}$ \\
$n_{\rm e}{\rm \;(OB)} \;(\pcm3)$ & $1.3 \times 10^{7}$ & $2.0 \times 10^{7}$ \\
$B \;{\rm (mG)}$ & 990 & 1200 \\
$M_{\rm A}$ & 11.4 & 11.9 \\
$\theta$ & $5^{\circ}$ & $35^{\circ}$ \\
\hline
\end{tabular}
\end{center}
\end{table}

\subsubsection{WR wind composition}
\label{sec:composition}
The thermal X-ray emission is largely from the shocked WR star wind
\citep{Pittard:2002b}. Hence, it is particularly sensitive to the
abundance of He and C since the free electrons are largely stripped
from these elements. For a pure He and C wind, the continuum flux
scales as $(2^{2} + ({\rm C/He})6^{2})$, where C/He is the ratio of
abundance by number of C and He \citep{Pollock:2005}.  Unfortunately,
the mass fraction of C in WC stars spans the range $0.2-0.6$, with no
strong sensitivity with spectral type. In this work, we assume that
C/He = 0.25 by number (C$ = 0.4$ by mass), as the ratio of
C\thinspace{\sc iv} 5471/He\thinspace{\sc ii} 5411 is comparable to
that in WR\thinspace90 \citep[][P.~Crowther,
priv. communication]{Dessart:2000}. We note that \citet{Eenens:1992}
had previously measured C/He = 0.15 for WR\thinspace140.  The O
abundance in WC stars is also poorly constrained, though the number
ratio of O/He is estimated as $0.03 \pm 0.01$ for WR\thinspace90 (WC7)
and WR\thinspace135 (WC8) \citep{Dessart:2000}. Since
\citet{Hillier:1999} had earlier obtained O/He=0.1 for WR\thinspace111
(WC5), we adopt ${\rm O/He}=0.05$ (${\rm O} = 0.1$ by mass). Our values for
C/He and O/He lie within the range considered by \cite{Pollock:2005},
while we note that in comparison \cite{Zhekov:2000} adopt C and O
abundances which are below the lower end of the ranges noted
above. With these assumptions, the mass fractions of hydrogen,
helium, and ``metals'' ($X$, $Y$, and $Z$ respectively) for the WC7
wind are $X=0.0$, $Y=0.5$, $Z=0.5$.  Solar abundances are assumed for
the O star (mass fractions $X=0.705$, $Y=0.275$, $Z=0.020$). Both
winds are assumed to have a temperature of $10,000\;{\rm K}$, with an
ionization structure of H$^{+}$, He$^{+}$, and CNO$^{2+}$.

\subsubsection{Collisionless shocks - electron thermalization, 
non-equilibrium ionization, and shock modification}
\label{sec:unmodified}
Previous models of the X-ray emission from the WCR in wide CWB systems
have usually assumed that the shocks are collisional, with immediate
equilibration of the post-shock electron and ion
temperatures. However, in reality they are collisionless
\citep[e.g.,][]{Draine:1993}, and there is growing observational
evidence that in such shocks the ratio of post-shock electron to ion
temperature, $T_{\rm e}/T_{\rm i}$, is a function of the shock speed
\citep[e.g.,][]{Rakowski:2005}. Temperature equilibration through
Coulomb collisions and plasma instabilities then occurs some distance
downstream. This process has been examined in WR\thinspace140 by
\citet{Zhekov:2000} and by \citet{Pollock:2005}, and can naturally
account for the observed soft X-ray continuum emission.  Since the
ionization timescale is of the same order as the timescale for energy
transfer from ions to electrons, it is not surprising that
observations of WR\thinspace140 also indicate that the post-shock
plasma has yet to reach ionization equilibrium \citep{Pollock:2005},
and that line-profile models which assume rapid ionization equilibrium
\citep{Henley:2003} are unable to reproduce the observed correlation
of X-ray line widths with ionization potential \citep{Henley:2005}.

Despite this progress, the physics
of the heating process in the shock layer remains poorly understood,
and shocks which are efficient at accelerating particles may
equilibrate their electron and ion thermal populations more rapidly, in
addition to reducing the equilibrium value of $T_{\rm e}$ and $T_{\rm
i}$ \citep{Ellison:2004}. In this case, the X-ray spectrum is softened
due to the reduction in the post-shock electron temperature behind a
weaker subshock (see Fig.~\ref{fig:shock_mod}). It is unclear whether,
and to what extent, shock modification may occur in WR\thinspace140,
but this mechanism could potentially play some role in creating
the low value of $T_{\rm e}$ observed.
A simple calculation reveals that even if particle accleration is
highly efficient, the non-thermal particles in the
precursor would have a negligible effect on the ionization of the pre-shock
gas, consistent with the observed lack of line emission near 
the wind terminal speeds \citep{Pollock:2005}.

A full understanding of WR\thinspace140's spectrum will eventually require 
a detailed hydrodynamical model which accounts for the large post-shock
equilibration timescales of the plasma ionization and electron and ion
temperatures, combined with the acceleration of non-thermal particles
and their effect on the shocks bounding the WCR and both the upstream
and downstream flow \citep[e.g.,][]{Pollock:2005}.  As the development
of such models is not trivial, here the synthetic X-ray spectra are
calculated from models of unmodified collisional shocks.  Given this
simplification and the uncertainties described above, we adopt a
rather pragmatic approach, and concentrate on obtaining a comparable
flux at $\sim 3$~keV. Absorption is negligible at this energy at the
chosen phase (see Fig.~\ref{fig:nt_modelB}) for reasonable viewing
angles, while at higher energies the relative softness of the observed
emission is clear evidence for $T_{\rm e} < T_{\rm i}$, with
the (less likely) possibility that shock modification also plays a
role (see Fig.~\ref{fig:xray2}).

The fact that the thermal X-ray luminosity from the WCR, $L_{\rm x}$,
scales as $\Mdot^{2}$ when the post-shock gas is largely adiabatic
means that even if shock modification occurs, and were to lead to
fairly substantial (e.g., factor of two) changes in $L_{\rm x}$,
mass-loss rates derived from models with unmodified shocks will be
much less affected. Moreover, the inferred mass-loss rates are largely
insensitive to the assumption of $T_{\rm e}=T_{\rm i}$. For instance,
setting $T_{\rm e}=0.5T_{\rm i}$ throughout the entire WCR
reduces the emission at 3~keV by 23\% compared to a model with 
$T_{\rm e}=T_{\rm i}$. An increase of only 14\% in the mass-loss rates is
required for the original flux to be regained. 
In this way we can still obtain useful estimates of the stellar
mass-loss rates from the modelling work which follows.

\subsubsection{WR mass-loss rate from thermal X-ray models}
\label{sec:xray_mdots}
Having specified the composition of the winds
(Sec.~\ref{sec:composition}) and unmodified, collisional shocks
(Sec.~\ref{sec:unmodified}), theoretical calculations of the X-ray
emission, based on axisymmetric 2D simulations of the stellar winds
and the WCR, can now be made.  The X-ray emissivity is calculated
using the MEKAL emission code \citep[][and references
therein]{Mewe:1995} for optically thin thermal plasma, and ionization
equilibrium and rapid thermalization of the electrons is assumed.

\begin{figure}
\begin{center}
\psfig{figure=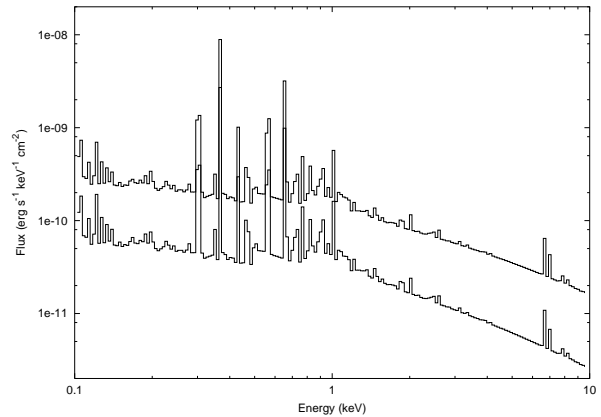,width=8.0cm}
\caption{Comparison of the intrinsic thermal X-ray emission calculated
from hydrodynamical models of the WCR in WR\thinspace140 at
$\phi=0.837$ for two different values of $\eta$, assuming unmodified
collisional shocks (Sec.~\ref{sec:unmodified}). $\eta=0.22$ (top) and
$\eta=0.0353$ (bottom).  In these calculations the mass-loss rate of
the WC7 star and the terminal wind speeds were fixed ($\Mdot_{\rm WR}
= 4.3 \times 10^{-5} \Msolpyr$, $v_{\infty, {\rm WR}} = 2860 \kmps$,
$v_{\infty, {\rm O}} = 3100 \kmps$), while the mass-loss rate of the O
star was varied according to the desired value of $\eta$. The
continuum flux is also sensitive to the adopted abundance of the WR
wind (see Sec.~\ref{sec:composition}).}
\label{fig:xray1}
\end{center}
\end{figure}

The intrinsic (i.e. prior to any absorption) X-ray emission calculated
from a model of the WCR with a fixed WC7 mass-loss rate, but for two
different values of $\eta$, is shown in Fig.~\ref{fig:xray1}. As
$\eta$ increases, the X-ray flux increases, as explained
earlier. Clearly, in order that the X-ray flux calculated from
hydrodynamical models of the WCR is comparable to observations, the
mass-loss rate of the WC7 star must be adjusted as $\eta$ is varied.

\begin{figure}
\begin{center}
\psfig{figure=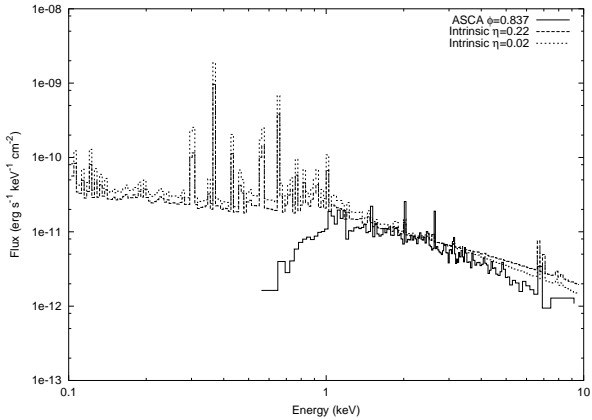,width=8.0cm}
\caption{The intrinsic thermal X-ray emission from hydrodynamical models of the
WCR at orbital phase $\phi=0.837$ with $\eta=0.22$ (dashed) and $\eta=0.02$
(dotted). There is a slight decrease in the hardness of the model spectra with 
$\eta$, as expected. Also shown is the observed X-ray emission (solid).
The X-ray emission from the models (at 3~keV) is comparable to the 
observed emission, though the latter is softer. As discussed in 
Sec.~\ref{sec:unmodified}, this is primarily because 
$T_{\rm e} < T_{\rm i}$.}
\label{fig:xray2}
\end{center}
\end{figure}

There have been a number of X-ray observations of WR\thinspace140 in
recent years, and we choose to use an ASCA observation taken on
Dec.~20~1999 ($\phi=0.837$). At this phase the X-ray luminosity is
climbing rapidly \citep{Pollock:2002}, while the radio emission is
around its maximum intensity, and the stellar separation,
\citep[$D=3.13\times10^{14}\cm$, see][]{Dougherty:2005}, is sufficiently
large that the shocked gas in the WCR is approximately adiabatic,
allowing easy adjustment of the model mass-loss rates in order to
obtain a specific X-ray flux.  The mass-loss rates required to match
the observed X-ray emission at $\phi=0.837$, as a function of $\eta$,
are noted in Table~\ref{tab:wr140_mdots}.  Fig.~\ref{fig:xray2}
demonstrates that the X-ray emission from two of these models is
comparable to the observed emission at $\sim 3$~keV, as
required. Models of the radio emission (Sec.~\ref{sec:radio_phi0.837})
indicate $\eta$ is likely to have a value near the lower end of the
range covered by Table~\ref{tab:wr140_mdots}. This implies the
mass-loss rate of the O4-5 supergiant in WR\thinspace140 is lower
($\approx 8\times10^{-7}\;\Msolpyr$) than the suggested values in
\citet{Repolust:2004} ($8.6-8.8\times10^{-6}\;\Msolpyr$). However, an
order of magnitude reduction in the mass-loss rate is consistent with
the recent estimates of \citet{Fullerton:2006}.

We note that \cite{Pollock:2005} determine $\Mdot_{\rm WR} \sim 1.5
\times 10^{-5} \Msolpyr$ from the degree of circumstellar absorption
observed by Chandra near periastron. While this value is independent
of the clumping in the winds, it is likely to be a lower limit since
the overall absorption of emission from the WCR is probably somewhat
less than occurs for emission near the apex of the WCR. This method
will also underestimate $\Mdot_{\rm WR}$ if there are regions of O
star wind between the observer and the WR star at this phase due to
the curvature of the WCR induced by the orbital motion of the stars.
Therefore, it is argued that the range of values of $\Mdot_{\rm WR}$
considered in our models is consistent with this constraint (see
Table~\ref{tab:wr140_mdots}), and does not preclude values of $\eta$
as low as 0.02.

\begin{table}
\begin{center}
\caption[]{The mass-loss rates (in units of $10^{-6} \Msolpyr$)
required to match the observed X-ray flux at 3~keV and $\phi=0.837$ as
a function of $\eta$. Column 4 indicates lower limits for the volume
filling factor, $f$, of clumps in the winds, as determined from a
comparison between the X-ray and radio derived mass-loss rates
(Sec.~\ref{sec:clump}). For simplicity, we assume that the two winds
have the same values of $f$. Recent analyses of O star winds indicate
that $f\sim 0.02-0.1$ \citep[see][and references
therein]{Fullerton:2006}. Columns 5-7 note approximate values of
$\theta$ for which the sight lines are parallel to the
asymptotic opening angles of the WR shock, contact discontinuity, and
O shock.}
\label{tab:wr140_mdots}
\begin{tabular}{lllllll}
\hline
\hline
$\eta$ & $\Mdot_{\rm WR} $ & $\Mdot_{\rm O}$ & $f$ & $\theta_{\rm WR}$ & $\theta_{\rm CD}$ & $\theta_{\rm O}$ \\
\hline
0.02 & 43.3 & 0.80 & 0.199 & 41$^\circ$ & 59$^\circ$ & 90$^\circ$ \\
0.0353 & 33.3 & 1.09 & 0.118 & 35$^\circ$ & 54$^\circ$ & 79$^\circ$ \\
0.055 & 27.2 & 1.38 & 0.079 & 30$^\circ$ & 49$^\circ$ & 71$^\circ$ \\
0.0825 & 23.2 & 1.76 & 0.057 & 25$^\circ$ & 42$^\circ$ & 65$^\circ$ \\
0.11 & 20.2 & 2.05 & 0.043 & 19$^\circ$ & 38$^\circ$ & 56$^\circ$ \\
0.16 & 17.2 & 2.54 & 0.031 & 11$^\circ$ & 32$^\circ$ & 52$^\circ$ \\
0.22 & 14.9 & 3.02 & 0.024 & 4$^\circ$ & 27$^\circ$ & 48$^\circ$ \\
\hline
\end{tabular}
\end{center}
\end{table}

Due to the relatively low abundance of C and O assumed by \cite{Zhekov:2000},
their derived mass-loss rates for $\eta=0.0353$ are higher than ours
(cf. their model A and AA), even though the system distance has since
been revised upwards. We also note that the wind parameters noted in
\cite{Dougherty:2005} overpredict the observed X-ray emission by a
factor of about 8. 

Given the existing uncertainties with respect to the degree of shock
modification, and the amount of post-shock electron heating and the
timescale for electron and ion temperatures to reach equilibrium,
Fig.~\ref{fig:xray2} demonstrates that it is difficult to uniquely
determine the WR star mass-loss rate and $\eta$ from the X-ray spectrum
alone. While a simpler, and perhaps a more reliable, method of
determining $\eta$ in CWB systems with highly eccentric orbits
involves utilizing the duration of any X-ray ``eclipse'' \citep[see,
e.g.,][]{Pittard:2002}, the very short eclipse seen in the
lightcurve of WR\thinspace140 \citep{Pollock:2005} means that a
detailed analysis using 3D hydrodynamical simulations is probably needed. 
Therefore, we turn to the high quality radio spectra now available
to further constrain $\eta$ (Sec.~\ref{sec:radio_phi0.837}).

\subsubsection{Wind clumping}
\label{sec:clump}
There is now considerable observational evidence that the winds of
massive stars are clumpy
\citep[e.g.,][]{Crowther:2002,Hillier:2003,Massa:2003,Repolust:2004,
Fullerton:2006,Owocki:2006}. Additional support is provided by models
of radiatively-driven winds. Clumping affects the thermal radio
emission from the unshocked winds, though simple arguments suggest
that in wide binaries, where the WCR is largely adiabatic, clumps are
rapidly destroyed upon encountering the WCR
\citep[][]{Pittard:2006}. Therefore, in the following models we assume
that the unshocked winds are clumpy (with the same degree of clumping
for both winds), and the WCR smooth.

The degree of clumping may be conveniently parameterized by a volume
filling factor, $f$, where it is assumed that the clumps occupy a
fraction $f$ of the wind volume, with the interclump medium being of
negligible density. Radio-derived mass-loss rates of clumpy winds are
related to those of smooth winds by $\Mdot = \Mdot_{\rm
smooth}/f^{1/2}$, since the thermal flux $S_{\rm ff}$ increases
accordingly \citep[see, e.g., Eq.~16 in][]{Pittard:2006}.  This
relationship is more complicated in binaries, as the degree of
clumping may be different for each wind, and the thermal emission from
the WCR may make a non-negligible contribution to the total thermal
emission \citep[see Fig.~11 in][]{Pittard:2006}. Nevertheless, we
proceed by assuming the WR wind dominates the thermal radio emission
in WR\thinspace140, so that a value for $f$ can be deduced from a
comparison of the X-ray and radio derived mass-loss rates. The
unclumped estimate of $5.3 \times 10^{-5} \Msolpyr$ from
\cite{Williams:1990} translates into $9.7 \times 10^{-5} \Msolpyr$ for
a system distance of 1.85~kpc, and leads to the lower limits for $f$
noted in Table~\ref{tab:wr140_mdots}.  An additional uncertainty is
that the mass-loss rate of \cite{Williams:1990} assumes the radio flux
at $\phi=0.0$ is entirely thermal, which is far from clear. While such
assumptions should be avoided if at all possible, our approach is
conservative in the sense that the theoretical thermal flux from the
unshocked winds cannot exceed the measured flux at periastron. Since
the total observed flux at $\phi=0.837$ is $\sim 20\times$ greater at
5~GHz and $\sim 5\times$ greater at 22~GHz than our estimate of the
thermal flux, the models in Sec.~\ref{sec:radio_phi0.837} should not
be overly sensitive to our choice of $f$, although $p$ will be
overestimated slightly if the thermal flux is in fact lower than
assumed. Wind clumping also affects the radius of the $\tau=1$ surface
\citep[see Eq.~17 in][]{Pittard:2006}.  In the clumpy WR wind in our
models, $R_{\tau=1}$ has values of 8.8, 4.0, and $1.8 \times 10^{14}
\;\cm$ at 1.6, 5, and 15~GHz. This is independent of the variation in
$\Mdot_{\rm WR}$ with $\eta$ due to the corresponding variation in
$f$.


\subsection{Models of the radio data}
\label{sec:wr140_radio}
We concentrate on obtaining reasonable spectral fits to the radio data
at the same orbital phase ($\phi=0.837$) we determined potential
values for stellar mass-loss rates and corresponding values of $\eta$
from the observed X-ray emission (see Sec.~\ref{sec:xray_mdots}).  The
brightness temperature of the radio emission is $\approx 3 \times
10^{7} \;$K at this phase, so the observed emission is dominated by
non-thermal processes \citep{Dougherty:2005}. Therefore, a good
estimate of the population of relativistic electrons can be obtained,
allowing predictions of the non-thermal X-ray and $\gamma$-ray
emission. We will apply our modelling to a variety of orbital phases
in a forthcoming paper.

There are several free parameters used to fit the observed radio
emission, including $\eta$, the index of the non-thermal electron
energy distribution, $p$, and $\zeta_{\rm rel,e}$ and $\zeta_{\rm B}$,
which normalize the model spectra and the magnetic field strength (and
hence the impact of the Razin effect). The effect of reducing the
luminosity of the stars is also investigated.  The line of sight angle
into the system at $\phi=0.837$ is specified by the orbit derivation
of \citet{Dougherty:2005} as $\theta=52^{\circ}$, though the
uncertainty in this value is large ($\sim 10^\circ$) and we allow
$\theta$ to vary substantially ($\pm 10-20^\circ$) in our models.  

\subsubsection{Parameter study}
\label{sec:param_space}
\begin{figure}
\begin{center}
\psfig{figure=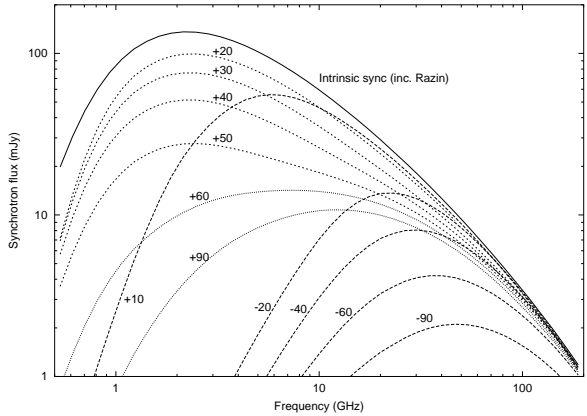,width=8.0cm}
\caption{The influence of $\theta$ on the observed {\em synchrotron} emission. 
All models have $\eta=0.11$, $p=2.0$, $\zeta_{\rm rel,e}=0.215$, and
$\zeta_{\rm B} = 2.6 \times 10^{-4}$. The lines of sight are through the WR
wind when $-90^{\circ} < \theta \ltsimm 19^{\circ}$, through the WCR
when $19^{\circ} \ltsimm \theta \ltsimm 56^{\circ}$, and through the O
wind when $56^{\circ} \ltsimm \theta < 90^{\circ}$, as indicated by
the long-dashed, short-dashed, and dotted lines respectively (see 
Table~\ref{tab:wr140_mdots}). The intrinsic synchrotron emission is indicated 
by the solid line.}
\label{fig:var_theta}
\end{center}
\end{figure}

Before attempting a fit to the $\phi=0.837$ data, it is instructive to
examine how the model spectra vary with $\theta$, $\eta$, $p$,
$\zeta_{\rm B}$, and $L_{\rm bol}$.  Fig.~\ref{fig:var_theta} shows
how the model {\em synchrotron} spectra change as $\theta$ (and hence the
free-free absorption along the line of sight) is varied,
for models with $\eta=0.11$, $p=2.0$, $\zeta_{\rm rel,e}=0.215$, and
$\zeta_{\rm B} = 2.6 \times 10^{-4}$. The Razin effect produces a
turnover at about 2.3~GHz, while the high frequency emission from each
model tends to the same slope as the free-free attenuation becomes
negligible. The sensitivity of the spectrum at low frequencies to
changes in $\theta$ is expected since the size of the radio
photosphere is larger than the stellar separation. At high frequencies
the radio photosphere is smaller and has a commensurately weaker
influence on the absorption of emission from the WCR. The synchrotron
emission is least affected by free-free absorption when $\theta
\approx 20^{\circ}$ (i.e. when the lines of sight are roughly parallel
to the asymptotic angle of the WR shock), and the turnover frequency
is the same as that of the intrinsic synchrotron emission. A dramatic
increase in the level of free-free absorption occurs as $\theta$
decreases and the WR wind moves in front of the WCR, with maximum
absorption at $\theta=-90^{\circ}$ (i.e. WR star in front). For
$\theta\ltsimm20^{\circ}$, free-free absorption causes the low
frequency turnover. As the line of sight swings through the WCR from
the WR shock to the O shock i.e. $\theta = 20\rightarrow56^{\circ}$,
the level of absorption again increases, although in a somewhat
complicated fashion as the source is extended and there are multiple
lines of sight.
While free-free absorption determines the slope of the spectrum
between 3 and $20-30$~GHz, it is only between $50^{\circ}$ and
$60^{\circ}$, when the sight lines move out of the WCR and into the O
wind, that the turnover due to free-free absorption dominates over the
Razin effect, and the turnover becomes very broad. The turnover
frequency increases as the O wind moves in front of the WCR i.e. as
$\theta \rightarrow 90^{\circ}$. The smaller stellar separation in the
models presented here results in free-free absorption having an
influence to much higher frequencies than in the models shown in
Fig.~10 of \citet{Dougherty:2003}.

It is interesting to note that when the sight lines are through the
WCR, a turndown or ``kink'' is visible in the spectrum at a few 10's
of GHz (this is most clearly seen when
$\theta=50^{\circ}$). Typically, such turndowns are attributed to a
change in the non-thermal energy spectrum. However, it is certainly
not the cause here, because the intrinsic synchrotron spectrum does
not display such a kink \citep[although spectral breaks from IC
cooling occur in our models, the break frequency is spatially
dependent, and is smoothed out once the flux is integrated over the
entire WCR - see Fig.~3 in][]{Pittard:2006}. It is clear that 
lines of sight through the WCR are optically thin at frequencies above 
20-30~GHz, and almost all, if not all, the synchrotron emission from
the WCR is seen. 

The variation of the spectra as a function of $\eta$ is shown in
Fig.~\ref{fig:var_eta}, where the total emission (synchrotron plus
free-free) is displayed. In this and the following figures, emission
between 0.5 and 50~GHz is examined, since this is approximately the
frequency range of the observations. The spectral variations in
Fig.~\ref{fig:var_eta} bear some resemblence to the changes with
$\theta$, since as $\eta$ varies, the relative angle of the sight
lines to the WCR changes if $\theta$ is fixed. For $\eta=0.055-0.16$,
the line of sight is through the lower opacity WCR. When $\eta=0.22$,
it is through the unshocked O wind, and the synchrotron emission
from the WCR suffers more free-free absorption than seen in the other
models, which results in the low frequency turnover occuring at a
higher frequency (c.f. Fig.~\ref{fig:var_theta}).  For the remaining
models, the reduced level of free-free absorption means that the
turnover is the result of the Razin effect, and therefore occurs at
approximately the same frequency in each case (since $\zeta_{\rm B}$
is fixed). It is clear from Fig.~\ref{fig:var_eta} that the spectral
slope between 10-40~GHz steepens as $\eta$ decreases (the slight
upward curvature in the $\eta=0.055$ model near 40~GHz is caused by
rising free-free emission), which is due to the change in the relative
angle of the sight lines with respect to the WCR, as seen in
Fig.~\ref{fig:var_theta}, and the resulting change in the free-free
opacity. While Fig.~\ref{fig:var_eta} does not show frequencies
as high as Fig.~\ref{fig:var_theta}, the synchrotron
emission tends to a similar slope in all models 
(since $p$ is the same in each model) with a slight steepening as 
$\eta$ decreases due to the WCR moving 
closer towards the O star, and the amount of IC cooling increasing.

Fig.~\ref{fig:var_p} shows the effect of varying the slope of the
non-thermal electron energy distribution. As $p$ decreases, the slope
of the intrinsic synchrotron emission increases, 
resulting in flatter attenuated spectra.
Fig.~\ref{fig:var_zetab} illustrates the effect of changing
$\zeta_{\rm B}$.  As $\zeta_{\rm B}$ is reduced, the spectral index
above 4~GHz increases. By decreasing $\zeta_{\rm B}$ (and thus $B$),
the synchrotron emission at a given frequency arises from non-thermal
electrons with higher Lorentz factors ($\nu \propto \gamma^{2}B$), and
thus sample a steeper slope for the non-thermal electron energy
distribution due to the increased impact of IC cooling at high
$\gamma$ \citep[see Fig.~3 in][]{Pittard:2006}. At very low values of
$\zeta_{\rm B}$, the Razin effect becomes the dominant mechanism
responsible for the low frequency turnover in
Fig.~\ref{fig:var_zetab}.

The effect of adjusting the stellar luminosities is shown in
Fig.~\ref{fig:var_lum}.  For simplicity, the luminosity of both stars
is adjusted by the same amount, although the amount of IC cooling is
primarily determined by the luminosity of the O supergiant. In
our model, a power-law distribution of non-thermal electrons is
injected along the entire shock front with the same initial spectral
index and normalization (relative to the local post-shock thermal
energy density - see Sec.~\ref{sec:norm_part}). 
Near the apex of the WCR, there is strong IC cooling but the
cooling weakens away from the apex as the stellar radiation field density
diminishes, and the non-thermal particles can maintain higher
energies. As the $L_{\rm bol}$ values are reduced, the influence of IC
cooling decreases, the non-thermal electrons are able to keep higher
energies throughout the WCR, and the population of the highest energy
electrons can exist throughout a larger part of the WCR.  The result
is that the spectrum of synchrotron emission, summed over the entire WCR, 
becomes less steep.  Clearly a specific spectral slope can be obtained with a
larger value of $p$ if the model values of $L_{\rm bol}$ are
reduced. For a given energy density of non-thermal electrons
immediately post-shock, a reduction in IC cooling also produces an
increase in the overall flux. It is conceivable that the assumptions
in Sec.~\ref{sec:wr140_parms} may overestimate the stellar
luminosities, but with one exception the models in the rest of this
work are based on the luminosities noted in
Table~\ref{tab:wr14067_params}.

\begin{figure}
\begin{center}
\psfig{figure=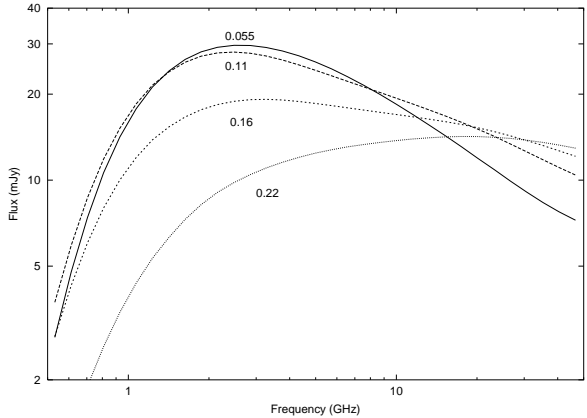,width=8.0cm}
\caption{The influence of $\eta$ on the model radio spectra. IC
cooling and the Razin effet are included in these calculations. Ionic
cooling and synchrotron self-absorption are negligible. All models
have $\theta=50^{\circ}$, $p=2.0$, $\zeta_{\rm rel,e}=0.215$, and
$\zeta_{\rm B} = 2.6 \times 10^{-4}$, though the mass-loss rates and
$f$ were varied according to Table~\ref{tab:wr140_mdots}. The line of
sight is through the WCR in all models except for $\eta=0.22$, when it
is through the unshocked O star wind.  The synchrotron emission from
the WCR dominates the free-free emission from the winds and WCR for
the chosen parameters.}
\label{fig:var_eta}
\end{center}
\end{figure}

\begin{figure}
\begin{center}
\psfig{figure=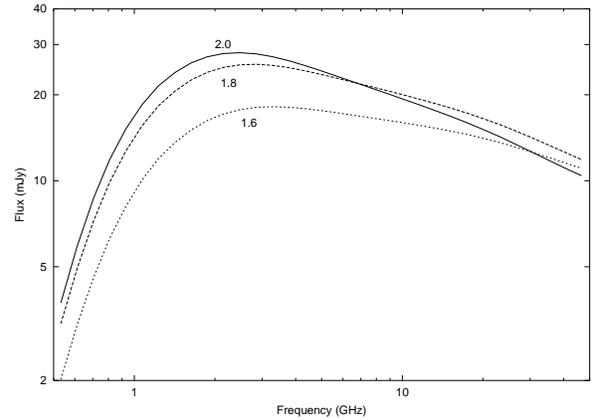,width=8.0cm}
\caption{The influence of $p$ on the model radio spectra. All models 
have $\eta=0.11$, $\theta=50^{\circ}$, $\zeta_{\rm rel,e}=0.215$, and
$\zeta_{\rm B} = 2.6 \times 10^{-4}$, and the line of sight is through the
WCR.}
\label{fig:var_p}
\end{center}
\end{figure}

\begin{figure}
\begin{center}
\psfig{figure=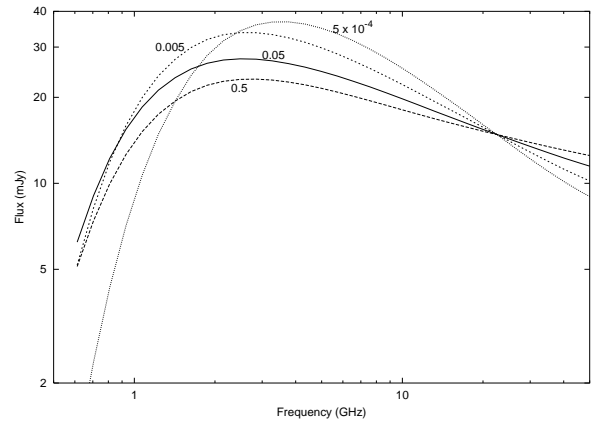,width=8.0cm}
\caption{The influence of $\zeta_{\rm B}$ on the model radio spectra. 
All models have $\eta=0.0353$, $p=1.4$, and $\theta=27^{\circ}$, and 
the line of sight is through the unshocked WR wind.
$\zeta_{\rm rel,e}$ was adjusted in each model to obtain a specific
flux at 22~GHz.}
\label{fig:var_zetab}
\end{center}
\end{figure}

\begin{figure}
\begin{center}
\psfig{figure=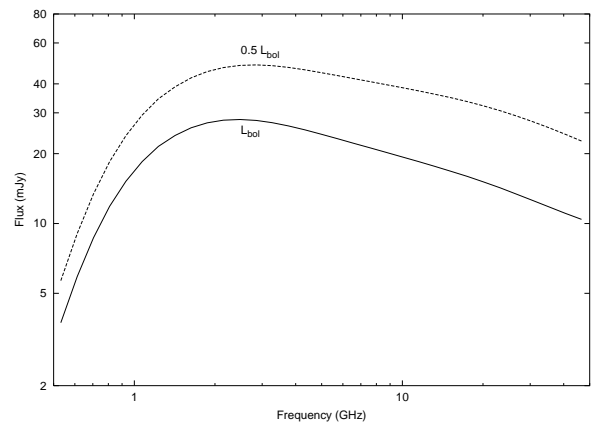,width=8.0cm}
\caption{The influence of the stellar luminosities on the model radio spectra. 
Both models had $\eta=0.11$, $p=2.0$, $\zeta_{\rm rel,e} = 0.215$,
$\zeta_{\rm B} = 2.6 \times 10^{-4}$, and $\theta=50^{\circ}$, and 
the line of sight is through the WCR.}
\label{fig:var_lum}
\end{center}
\end{figure}

We have also investigated what happens to the observed
synchrotron emission if the emission region is limited to $r < \pi
r_{\rm O}/2$, as suggested by \citet{Eichler:1993}. We find that the
spectral index is essentially unchanged, and thus our fits for $p$ are 
robust in this context, though the flux is diminished to
$40-75$\% of its previous value due to the reduced size of the
emitting region.

\subsubsection{Spectral fits at $\phi=0.837$}
\label{sec:radio_phi0.837}
Our inital aim was to determine parameter values directly from model
fits to the radio data. Unfortunately, reasonable fits to the data are
possible with a range of parameters. For example, by varying $p$ and
$\theta$ similar fits are attained over a wide range of
$\eta$. Furthermore, the data can be fit with either free-free
absorption or the Razin effect responsible for the low frequency
turnover. However, despite this degeneracy of models, it is possible
to discriminate between models by using other properties, such as the
values of $\theta$ and $\zeta_{\rm rel,i}$ required by each model,
which we discuss later in this section. Parameters for a variety of
model fits are listed in Table~\ref{tab:goodfits}.

\begin{figure}
\begin{center}
\psfig{figure=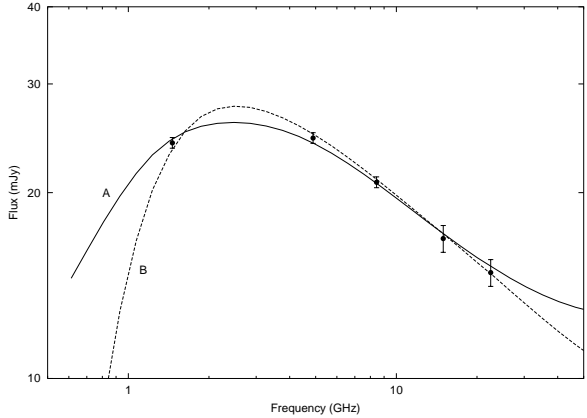,width=8.0cm}
\caption{Model fits to radio data of WR\thinspace140 at $\phi=0.837$.
In these models, which span the range of $\eta$ considered, the low
frequency turnover is due to free-free absorption. The data are
represented by solid circles, with fluxes taken from Table~2 of 
\citet{Dougherty:2005}; in this and subsequent figures we assume
that the uncertainty in the absolute flux scale is $5\%$ for $\nu \geq
15$~GHz and $2\%$ for $\nu < 15$~GHz \citep{Perley:2003}.  The model
parameters are noted in Table~\ref{tab:goodfits}, and the line of
sight is through the unshocked WR wind. Model~A has $\eta=0.22$, while
model~B has $\eta=0.02$.  The low frequency turnover is shallower in
model~A, and the high frequency total emission has a stronger concave
curvature, relative to model~B. The variation of the high frequency
emission is due to differences in the slope of the synchrotron and
free-free emission in both models. The thermal flux from the unshocked
winds and the WCR is $\approx 4$~mJy at 50~GHz in both models.}
\label{fig:ffabs_fits}
\end{center}
\end{figure}

Two model fits which span the range of $\eta$ considered are shown in
Fig.~\ref{fig:ffabs_fits}. For both of the models in
Fig.~\ref{fig:ffabs_fits}, $\zeta_{\rm B}=0.05$ was assumed.
Sightlines which have $\theta$ marginally less than $\theta_{\rm WR}$
are necessary for free-free absorption to cause the low frequency
turnover.  With these restrictions, $p=1.4$ is needed to match the
slope of the data points between 5 and 22~GHz. Fits with free-free
absorption dominant over Razin cannot be obtained when the sightlines
are through the WCR because the free-free absorption is too low, while
sightlines through the unshocked O wind require $p\gtsimm2.5$ for the
non-thermal electron distribution to offset the synchrotron spectrum
between 5-22~GHz becoming flatter as $\theta \rightarrow 90^{\circ}$
and the free-free opacity increases
(cf. Fig.~\ref{fig:var_theta}). While such values of $p$ are
appropriate for modified shocks, the required values of $\theta$ are
much too high compared to the value of $52^{\circ}$ determined by the
orbital solution of \citet{Dougherty:2005}. In fact, we find that all
model fits with $p > 2$ are unsatisfactory, with either $\zeta_{\rm
rel,e}$ or $\theta$ being uncomfortably high.

Since the opening angle of the WCR varies with $\eta$, the value of
$\theta$ required to match the low frequency synchrotron turnover
varies in a similar fashion. For the models in
Fig.~\ref{fig:ffabs_fits}, $\theta$ is $5^{\circ}$ for model~A and
$35^{\circ}$ for model~B (acceptable fits with $\eta$ between 0.02 and
0.22 can also be obtained). The value of $\theta$ from model~A is
inconsistent with the $52^{\circ}$ determined by \cite{Dougherty:2005}
at this phase, and while in model~B it is still different by some
considerable margin, it is not entirely implausible given the large
uncertainty that exists for $\theta$. It is anticipated that models
with $\eta < 0.02$ should result in fits with $\theta$ closer to the
value determined by \cite{Dougherty:2005}, but $\eta$ is not expected
to be lower than $\sim 0.01$ because the mass-loss rate of the WR star
then becomes too large.

\begin{figure}
\begin{center}
\psfig{figure=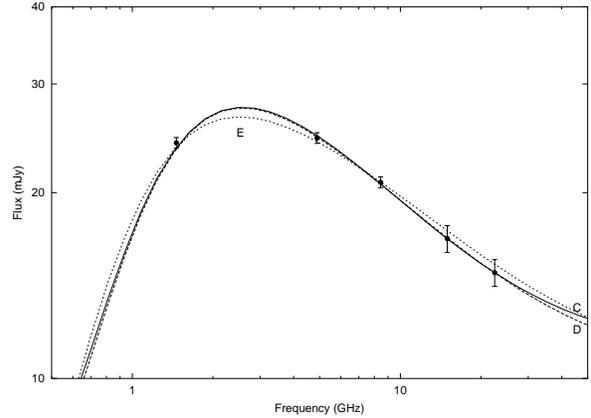,width=8.0cm}
\caption{Model fits to the radio data at $\phi=0.837$, again where the
low frequency turnover is due to free-free absorption (models~C, D,
and~E).  Due to the sensitivity of the slope of the high frequency
synchrotron emission to both $p$ and $\zeta_{\rm B}$, it is possible
to obtain, for a given $\eta$, fits of similar quality with different
values of $p$ and $B$.  All models had $\eta=0.11$ (see
Table~\ref{tab:goodfits}).}
\label{fig:families}
\end{center}
\end{figure}

\begin{figure}
\begin{center}
\psfig{figure=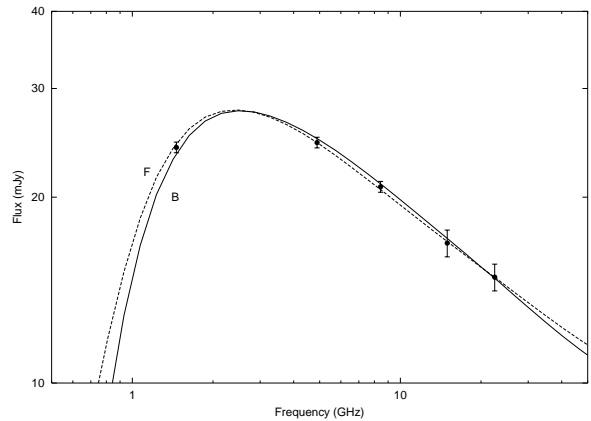,width=8.0cm}
\caption{Models of the radio data at $\phi=0.837$ which
demonstrate that it is possible to obtain fits of similar quality
with a higher value of $p$ when the stellar luminosities are
reduced. Both models (B and F) had $\eta=0.02$ and $\zeta_{\rm B}=0.05$ 
(see Table~\ref{tab:goodfits}).}
\label{fig:lbolfits}
\end{center}
\end{figure}

The slope of the high frequency synchrotron emission is sensitive to
both $p$ (Fig.~\ref{fig:var_p}) and $\zeta_{\rm B}$
(Fig.~\ref{fig:var_zetab}). Since there is almost complete degeneracy
between these parameters, a variety of fits with different families of
$p$ and $\zeta_{\rm B}$ can be obtained (Fig.~\ref{fig:families}). For
a given $\eta$, the value of $\theta$ required to match the low
frequency turnover is largely unchanged in these fits.  If the stellar
luminosity is reduced, fits with a higher value of $p$ can be
obtained, as shown in Fig.~\ref{fig:lbolfits}. However, we are unable
to fit the data with $\eta=0.02$ and $p=2$, despite making the other
parameters as favorable as possible for this task (\eg, $\zeta_{\rm
B}=0.05$ and luminosities reduced by a factor of two).  

\begin{figure}
\begin{center}
\psfig{figure=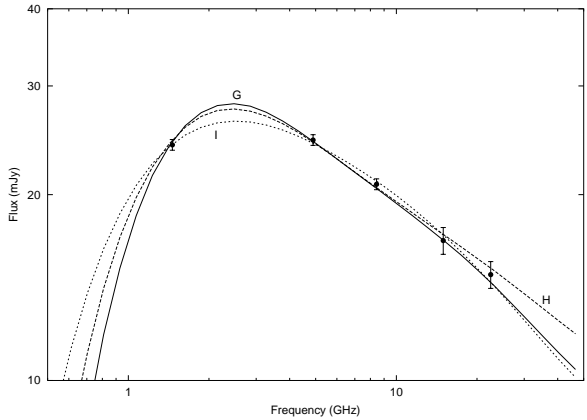,width=8.0cm}
\caption{Model fits to the radio data at $\phi=0.837$ where the 
low frequency turnover is due to the Razin effect (models~G, H, and~I). 
The model parameters are again noted in Table~\ref{tab:goodfits}.}
\label{fig:razin}
\end{center}
\end{figure}

As already noted, we are also able to obtain reasonable fits to the
data when the Razin effect is responsible for the low frequency
turnover.  Again, fits of similar quality can be obtained across a
wide range of $\eta$ and $p$, as demonstrated in Fig.~\ref{fig:razin}.
In these models an attempt was made to match $\theta$ as closely as
possible to the value determined from the orbital solution of
\cite{Dougherty:2005}.

\begin{figure}
\begin{center}
\psfig{figure=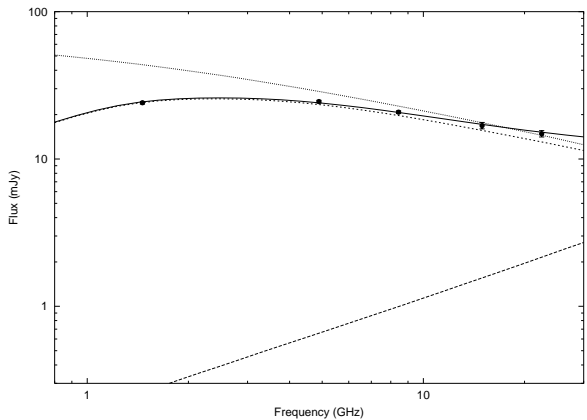,width=8.0cm}
\caption{The amount of free-free absorption affecting the intrinsic
synchrotron emission in model~A, where the low frequency turnover is
due to free-free absorption.  Various emission components are shown -
free-free flux (long-dashed), synchrotron flux (short-dashed),
intrinsic synchrotron flux (dotted), and total flux (solid).
The spectral index of the intrinsic synchrotron emission is $-0.34$
between 1.5 and 5~GHz, and steepens to $-0.47$ between 15 and
22.5~GHz. This illustrates the effect of IC cooling, since, for $p=1.4$,
$\alpha=-0.2$ is expected.}
\label{fig:ffabs_components}
\end{center}
\end{figure}

\begin{figure}
\begin{center}
\psfig{figure=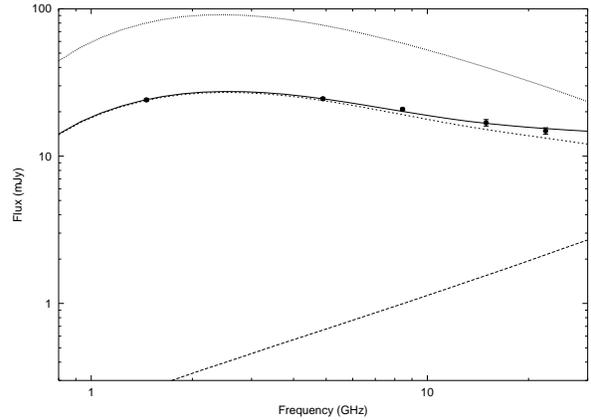,width=8.0cm}
\caption{As Fig.~\ref{fig:ffabs_components} but for model~J, where the 
Razin effect is responsible for the low frequency turnover.}
\label{fig:razin_components}
\end{center}
\end{figure}

\begin{table}
\begin{center}
\caption[]{Parameters for model fits to the $\phi=0.837$ radio data. In 
models A-F the low frequency turnover is due to free-free absorption, 
while in models G-J it is due to the Razin effect. The values of
$\theta$ for which the sight lines are parallel to the asymptotic
opening angle of the WR shock, contact discontinuity, and O shock are
noted in Table~\ref{tab:wr140_mdots}.}
\label{tab:goodfits}
\begin{tabular}{llllll}
\hline
\hline
Model & $\eta$ & $p$ & $\zeta_{\rm rel,e}$ & $\zeta_{\rm B}$ & $\theta$\\
\hline
A & 0.22 & 1.4 & $1.38 \times 10^{-3}$ & 0.05 & $5^{\circ}$ \\
B & 0.02 & 1.4 & $5.36 \times 10^{-3}$ & 0.05 & $35^{\circ}$ \\
C & 0.11 & 1.53 & $2.26 \times 10^{-4}$ & 0.5 & $13^{\circ}$ \\
D & 0.11 & 1.4 & $2.03 \times 10^{-3}$ & 0.05 & $13^{\circ}$ \\
E & 0.11 & 1.1 & $2.79 \times 10^{-2}$ & $5.0 \times 10^{-3}$ & $14^{\circ}$ \\
F & 0.02 & 1.6 & $1.81 \times 10^{-3}$ & 0.05 & $40^{\circ}$ \\ 
G & 0.11 & 2.0 & 0.22 & $2.6 \times 10^{-4}$ & $50^{\circ}$ \\
H & 0.11 & 1.6 & 0.15 & $4.0 \times 10^{-4}$ & $44^{\circ}$ \\
I & 0.0353 & 1.4 & 0.14 & $1.0 \times 10^{-3}$ & $58^{\circ}$ \\
J & 0.22 & 1.5 & 0.14 & $4.0 \times 10^{-4}$ & $36^{\circ}$ \\
\hline
\end{tabular}
\end{center}
\end{table}

Before the relative merits of the models with either the Razin effect
or free-free absorption being responsible for the low frequency
turnover are considered, it is worth looking at the make-up of each of
these models.  The intrinsic synchrotron (i.e. before circumstellar
absorption) and the free-free emission components to the total flux
for model~A where the low frequency turnover is due to free-free
absorption are shown in Fig.~\ref{fig:ffabs_components}.  The
free-free flux is negligible in comparison to the synchrotron flux,
which only suffers significant absorption at low frequencies.  In this
case, the observed synchrotron luminosity between 1-40~GHz is roughly
90\% of the intrinsic synchrotron luminosity.  Somewhat paradoxically,
the intrinsic synchrotron emission from models where the low frequency
turnover is caused by the Razin effect can experience more severe
free-free absorption, as seen from a comparison of
Figs.~\ref{fig:ffabs_components} and~\ref{fig:razin_components}.  This
is because the free-free absorption is smallest when the sight lines
are closely parallel to the WR shock (as is the case for model~A), and
is significantly greater when the sight lines are deeper within the
WCR \citep[e.g., compare the 10~GHz flux in Fig.~\ref{fig:var_theta}
when $\theta=10$ or $20^{\circ}$ with the flux when
$\theta=40^{\circ}$ - see also the dotted curve in Fig.~12b
in][]{Pittard:2006}. The observed synchrotron luminosity in
Fig.~\ref{fig:razin_components} is only 40 per cent of the intrinsic
synchrotron luminosity, and in such cases the IC luminosity predicted
using Eq.~\ref{eq:icsynclum} will be substantially underestimated if
the absorption is not accounted for.
 
\begin{figure*}
\begin{center}
\psfig{figure=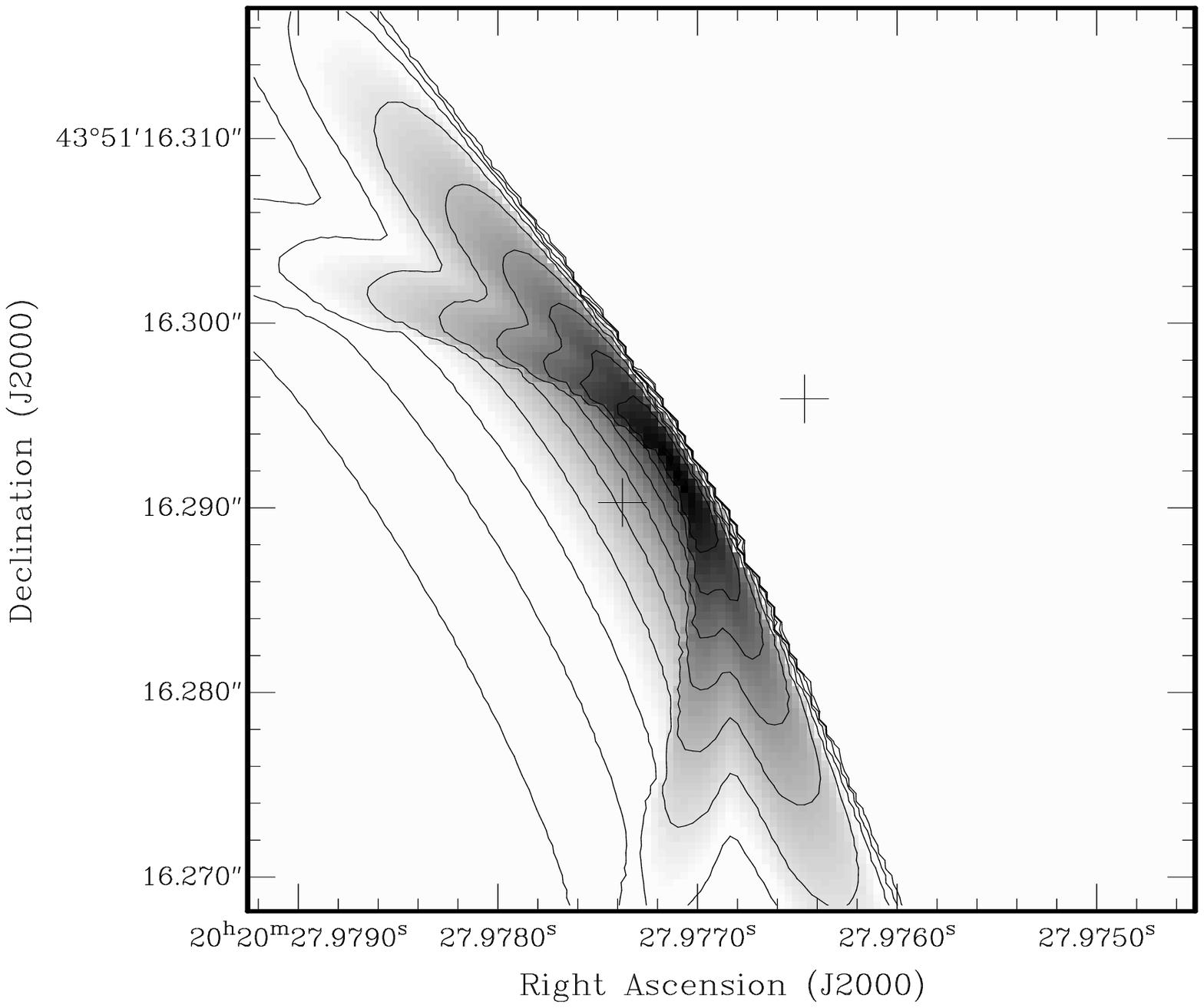,width=8.0cm}
\psfig{figure=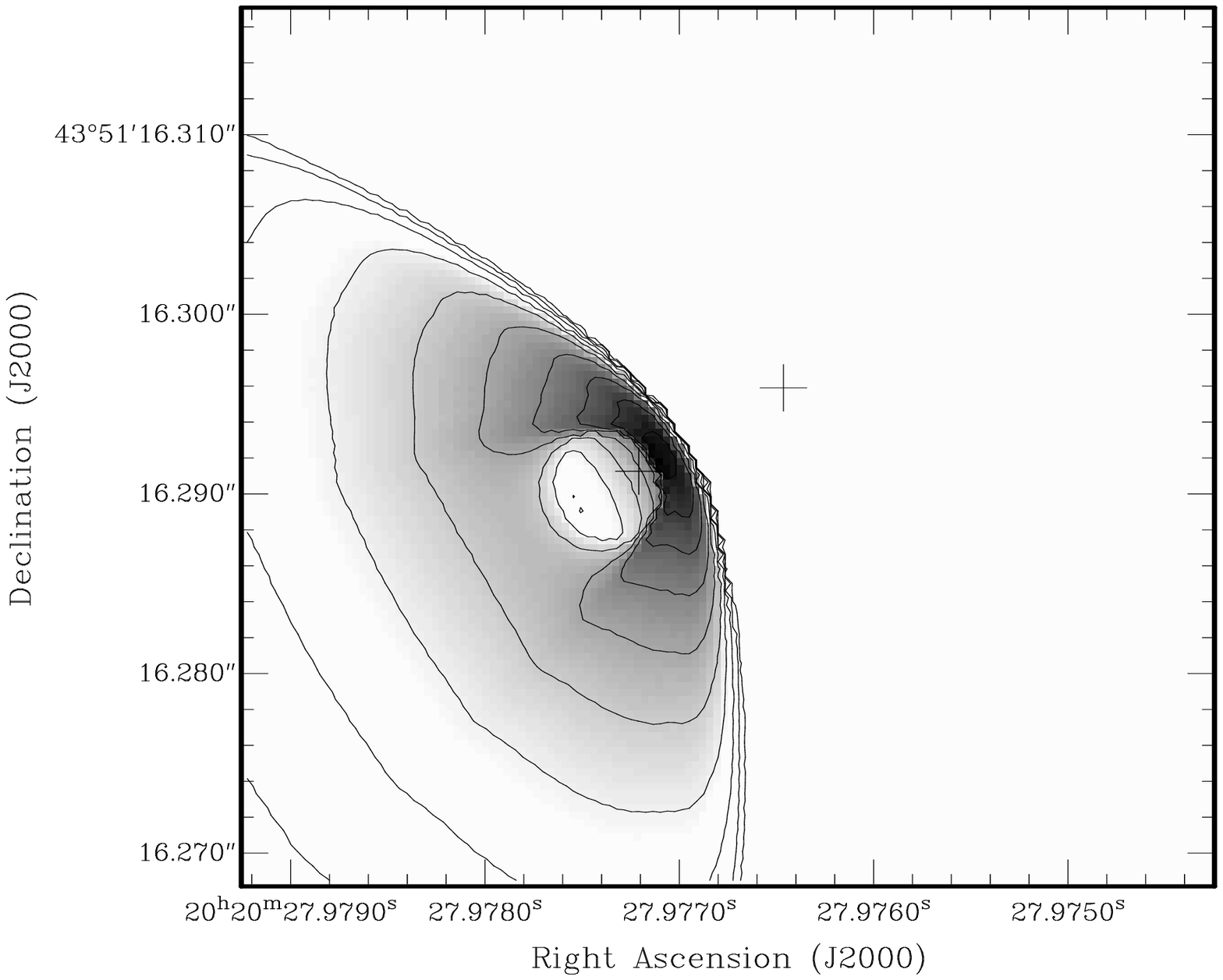,width=8.0cm}
\psfig{figure=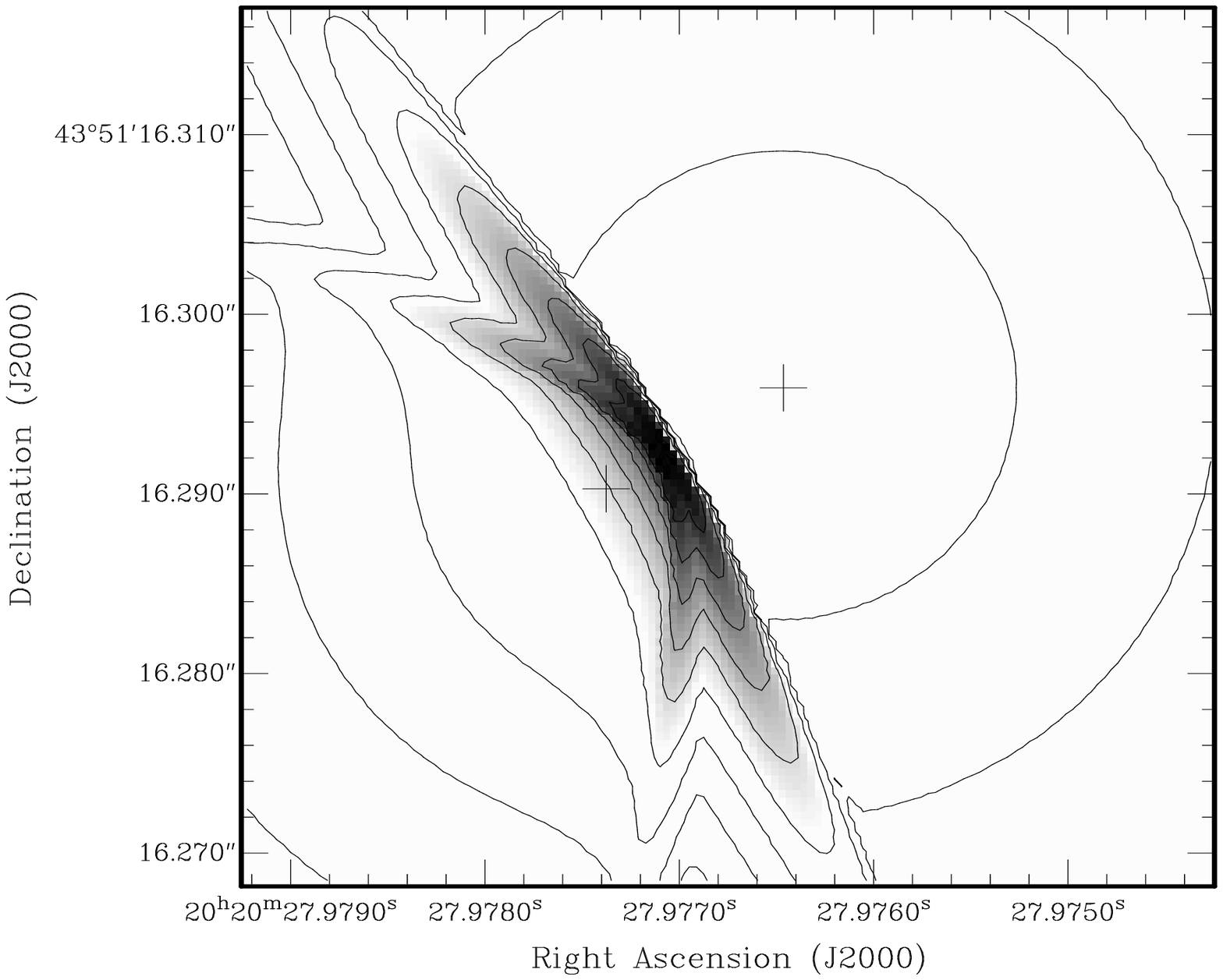,width=8.0cm}
\psfig{figure=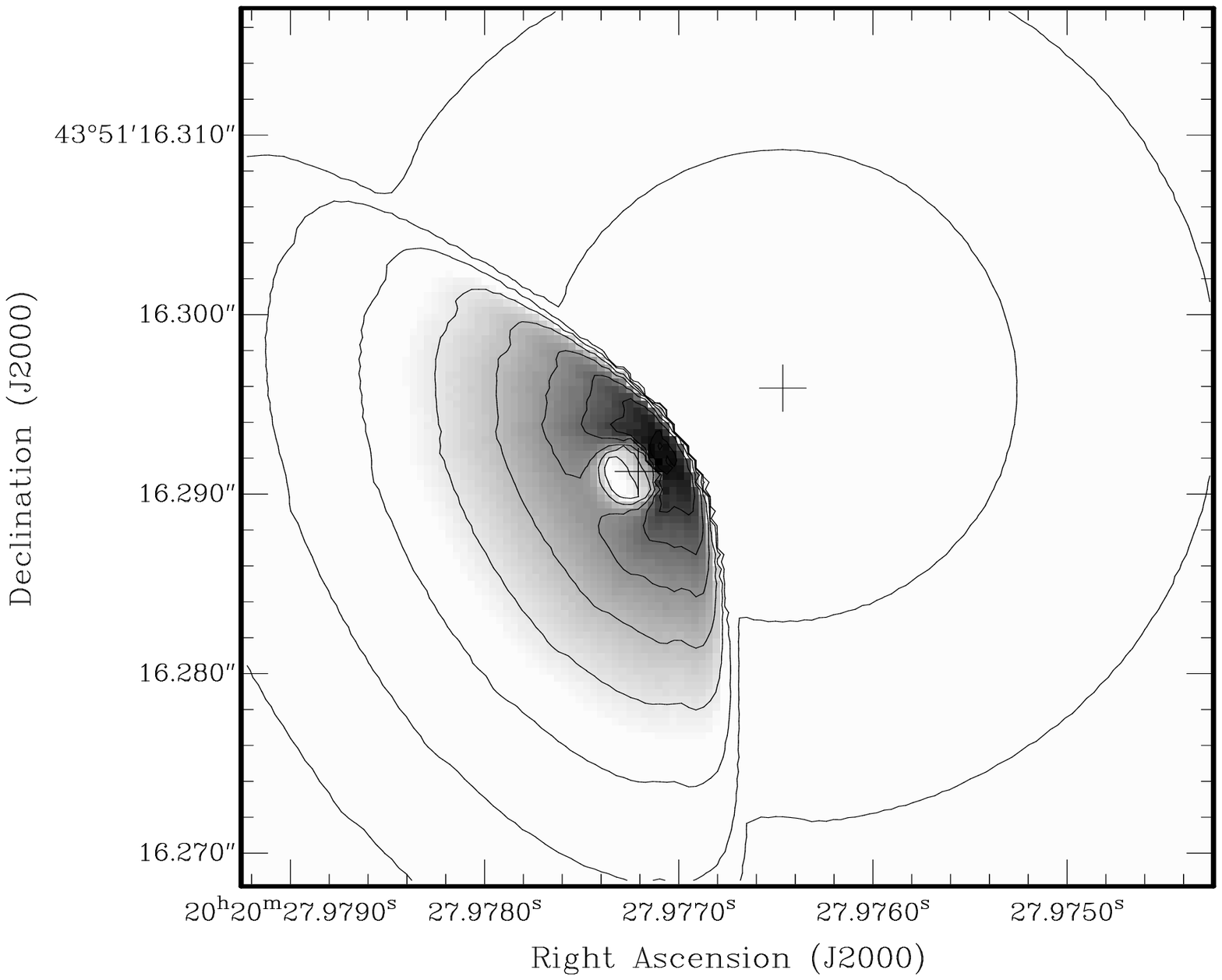,width=8.0cm}
\caption[]{Intensity distributions at 1.7~GHz (top) and 8.4~GHz (bottom) 
from model~A (left) and model~B
(right) of WR\thinspace140. The contour levels are spaced apart by
0.5~dex, are chosen to highlight the large-scale structure, and are 
identical in both model's at a specific frequency, though change with 
frequency. The positions of the stars are indicated by crosses (WR star to the
upper-right of the O star). The effect of different wind momentum ratios and
line of sight angles are clearly visible.}
\label{fig:wr140images}
\end{center}
\end{figure*}

\subsubsection{Synthetic radio images}
\label{sec:radio_images}
The spatial distributions of radio emission from models A and B are shown
in the left and right panels of Fig.~\ref{fig:wr140images}. In all 
panels the intensity of emission from the WCR exceeds that from the stars, 
though the relative amount of free-free emission increases strongly
with frequency. Model~A is viewed at low inclination, while model~B is
viewed at a somewhat higher inclination (Table~\ref{tab:goodfits}).
The absorption of emission from the far side of the WCR by the O wind
is negligible in model~A, but significant in model~B, as evident by
the dearth of emission near the O star.  As expected, this hole is
much larger at 1.7~GHz than at 8.4~GHz. The O star is not located at
the centroid of the hole, since the O wind is restricted on one side
by the WCR. Those parts of the WCR with optically thin lines of sight
through the outer envelope of the ``modified'' O wind are
predominantly observed. In addition, the intensity of the far side of
the WCR is highest nearer the apex.  The reduced opening angle of the
WCR in model~B relative to model~A is clearly visible. IC cooling
creates a ``V''-shaped intensity distribution in model~A, but this is
not so clearly seen in model~B, largely due to the difference in
viewing angle.

Though not shown here, another important finding is that the relative
position of the stars and the observed peak of the emission from the
WCR is not necessarily related to $\eta$, due to the occultation of
the WCR by the O star stellar wind.  In fact, as $\theta$ increases
and $D$ decreases the absorption can become so large that the position
of the WCR apex appears closer to the WR star than to the O star,
despite the WR star having the stronger wind. 
Finally, we note that no offset between the center of
emission at 1.7~GHz and 8.4~GHz is seen, contrary to the report by 
\cite{Dougherty:2005}.

\begin{figure}
\begin{center}
\psfig{figure=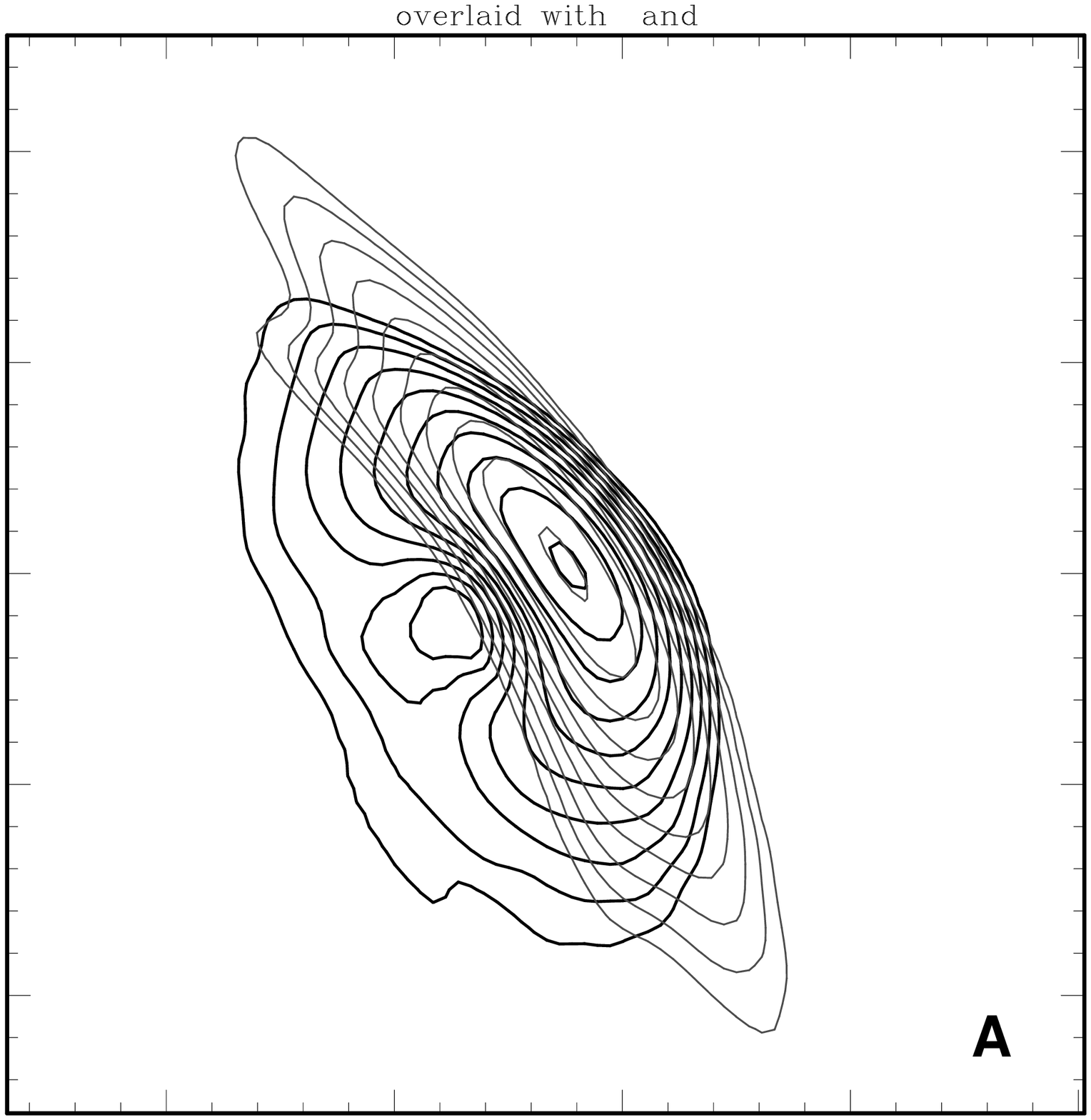,clip=,bbllx=26pt,bblly=150pt,bburx=569pt,bbury=693pt,width=5.9cm}
\psfig{figure=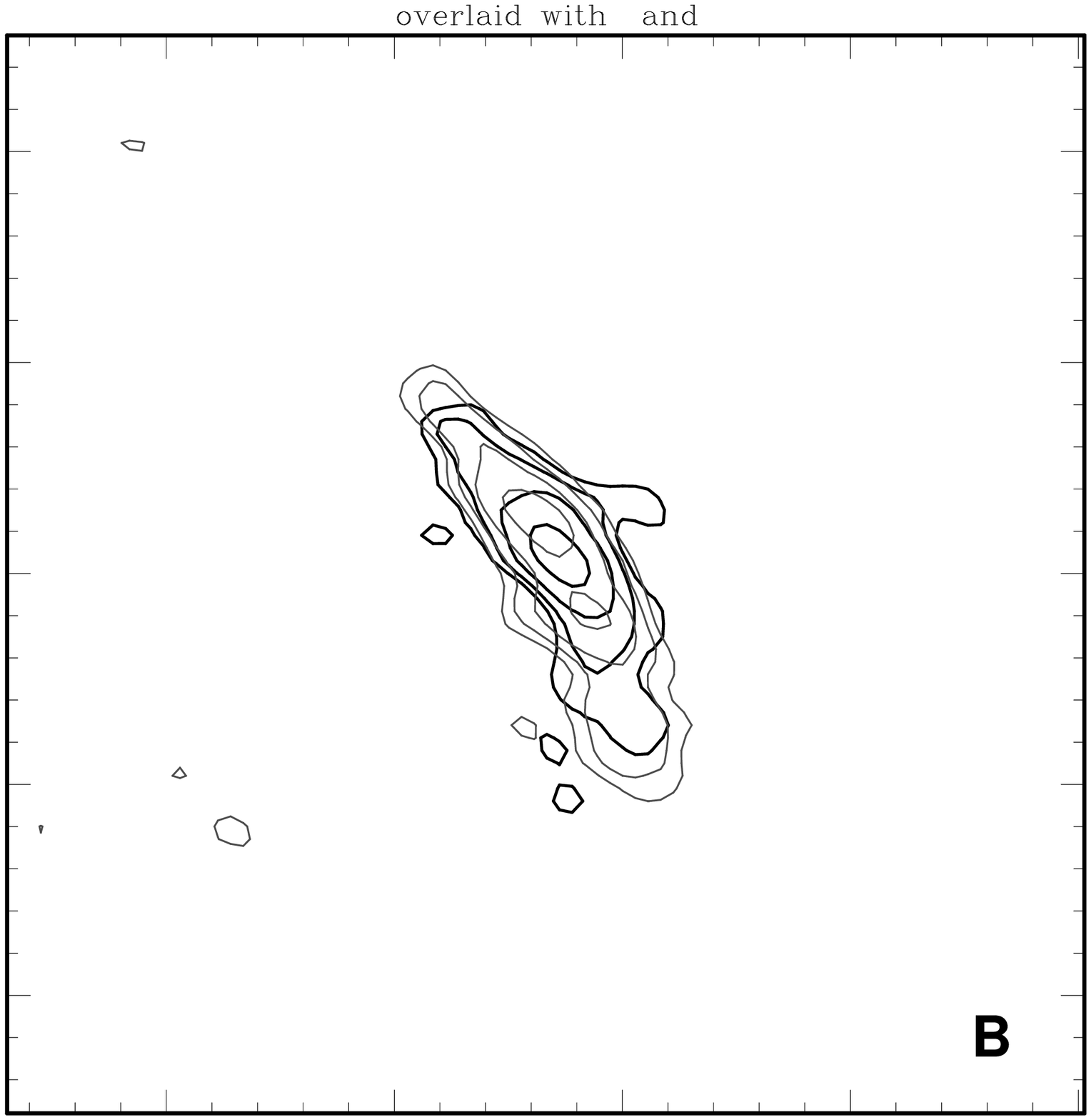,clip=,bbllx=26pt,bblly=150pt,bburx=569pt,bbury=693pt,width=5.9cm}
\psfig{figure=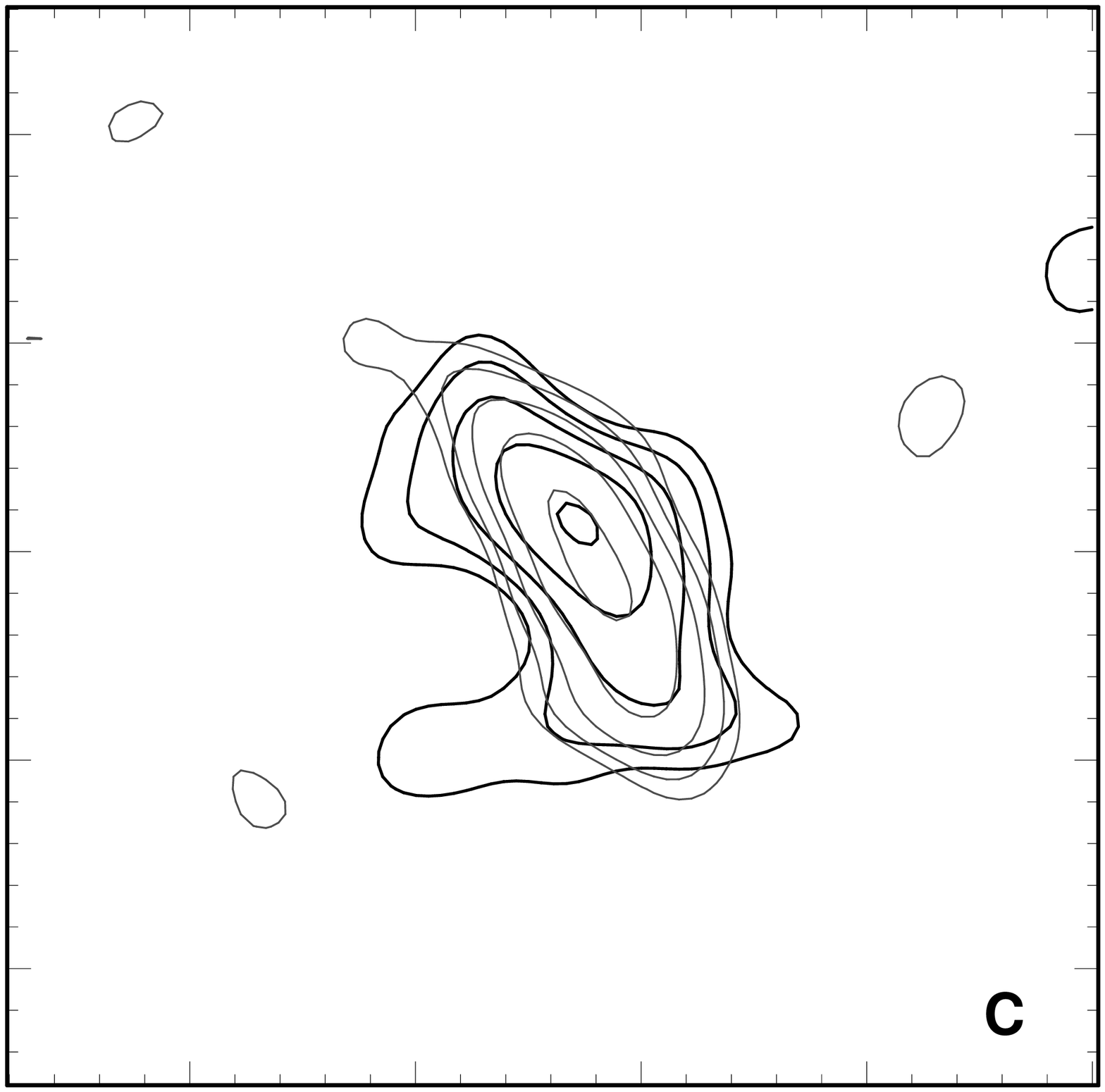,clip=,bbllx=26pt,bblly=150pt,bburx=569pt,bbury=693pt,width=5.9cm}
\caption[]{8.4 GHz synthetic VLBA observations of the model intensity
distributions in Fig.~\ref{fig:wr140images}. 
The data have been scaled to match the peak flux in the observations
in \citet{Dougherty:2005} at phase 0.841 - close to 1.2~mJy$\;{\rm
beam}^{-1}$. The top panel (a) shows the resulting image for model A
(thin) and model B (thick) when there is no noise i.e. ``perfect''
data. The middle panel (b) used the same imaging and deconvolution
parameters as in \citet{Dougherty:2005}, and shows no discernable
difference in the images of the two models. In the lower panel (c),
the lowest spatial frequency content of the data is more strongly
emphasized by severely tapering the data, and the two models may just
be distinguished. Contours in (a) start at $30\;\mu$Jy$\;{\rm
beam}^{-1}$ and increase by powers of 1.5. For (b) and (c), the first
contour is at $33\;\mu$Jy$\;{\rm beam}^{-1}$. Each minor tick on the
axes represents 1~mas. The scale and distribution of the emission in
panel (b) is consistent with the observations at both phase 0.830 and
0.841 presented in \citet{Dougherty:2005}.}
\label{fig:wr140_fake_images}
\end{center}
\end{figure}

To more closely compare the model intensity distributions at 8.4~GHz
to the VLBA images shown in \citet{Dougherty:2005}, we have used the
AIPS subroutine UVCON to generate visibilities appropriate for a VLBA
``observation'' of our models. System noise estimates were accounted
for by including the performance characteristics of the VLBA
telescopes, e.g., antenna efficiencies, system temperatures etc. The
resulting visibilities are then imaged and deconvolved using standard
techniques. The resulting simulated observations are shown in
Fig.~\ref{fig:wr140_fake_images}.  With perfect data i.e. no noise, it
is possible to distinguish the two different models
(Fig.~\ref{fig:wr140_fake_images}a). However, once appropriate noise
has been incorporated, the situation is quite different as can be seen
in Fig.~\ref{fig:wr140_fake_images}b and c. Adopting the same imaging
parameters (imaging weights etc.) as used in \citet{Dougherty:2005},
the differences between the models are not apparent. However,
applying a more extreme taper to strongly emphasise the lower
spatial frequency content of the visibility data
(Fig.~\ref{fig:wr140_fake_images}c) the different models can just be
discerned by an apparent difference in the opening angle. These
synthetic images show that any curvature in the WCR, from either
free-free absorption or from the wind-momentum ratio, is at best
difficult, and may be impossible, to detect with VLBI observations. It
is clearly dependent on the spatial frequency content and
the attainable signal-to-noise ratio of the data. The
high-brightness emission that is detected by VLBI arrays arises
predominantly in the vicinity of the shock apex, where the opening
angle of the shocks have yet to reach their asymptotic values. Even
when the lower surface brighness emission from electrons further down
the post-shock flow can be detected, e.g., by arrays with shorter
baselines to recover lower spatial frequency visibilities, the beam of
the array smooths out the features of the WCR. This leads us to call
into question the reliability of values of $\eta$ derived from these
types of observation.


\subsubsection{Discussion of the radio models}
\label{sec:eta_comp}
In attempting to identify the most suitable models, it is noted that
the fits in Fig.~\ref{fig:razin} require a low B-field within the WCR
($\zeta_{\rm B} \sim 3 \times 10^{-4} - 10^{-3}$), which is far from
equipartition with the energy density of non-thermal particles
(electrons plus ions). Assuming a simple dipole field, this implies
that the B-field at the surface of each star is very low, with little
or no amplification in the WCR. As the synchrotron flux is
proportional to $\zeta_{\rm rel,e} \;\zeta_{\rm B}^{3/4}$, large
values of $\zeta_{\rm rel,e}$ are required to normalize the model
spectra to the data. The models (G-J) shown in Figs.~\ref{fig:razin}
and~\ref{fig:razin_components} imply that 10-20 \% of the available
wind kinetic power processed through the shocks bounding the WCR
(which is a small fraction of the total kinetic power of the winds in the
system) is transferred to the non-thermal {\em electrons}. This is
unacceptably high. Hence, models where the low frequency turnover is
due to free-free absorption are preferred. The high energy non-thermal
emission from models G-J is also less consistent with current EGRET
and INTEGRAL observations, and in model~G the non-thermal X-ray flux
exceeds the observed thermal flux. This is discussed further in
Sec.~\ref{sec:high_en_em}.

Amongst models A-F, model~E is notable for an extremely low value of
$p$. The resulting high energy non-thermal spectrum is too flat if
WR\thinspace140 is indeed associated with the EGRET source
3EG~J2022+4317 (see Sec.~\ref{sec:nt_comp_obs}). The magnetic energy
density in model~C is towards the high end of expectations, with
$B=3.3$~G at the apex of the WCR. A magnetic field this high is likely
to prevent particle acceleration as the shock velocity does not exceed
the phase velocity of whistler waves propagating normal to the shock
\citep[see][for the necessary conditions]{Eichler:1993}. At this stage
we also prefer to use the stellar luminosities determined in
\citet{Dougherty:2005}, rather than reduced values such as used in
model~F, though further investigation of this matter is needed. On
this basis, models A, B and D are preferred. Of these three models,
the viewing angle of model~B is closer to the value determined by
\citet{Dougherty:2005}, and hence is the preferred model overall.

We now compare our estimate of $\eta$ from model~B to previous
estimates. First, $\eta=0.02$ is consistent with the value derived
from \citet{Marchenko:2003}, from an analysis of the C{\sc III} and
He{\sc I} line profiles around periastron. \citet{Dougherty:2005}
estimate a value an order of magnitude higher (0.22), but this is an
indirect estimate based on an assumed mass-loss rate for the O
supergiant. They also argue that $\eta=0.22$ is consistent with the
opening angle of the WCR as estimated from the VLBA images. However,
as argued in the previous section, we should not expect the opening
angle of the WCR to be constrained by the VLBA observations at this
orbital phase.  \citet{Varricatt:2004}, from He{\sc I} absorption in
the WR wind, determine a lower limit to the half-opening angle of
$42^{\circ}$, which is consistent with the value for the WR shock in
model~B.

\subsection{The high energy emission}
\label{sec:high_en_em}

\subsubsection{Models of the high energy emission}
\label{sec:calc_em}
The parameters obtained from the models of the radio data in 
Sec.~\ref{sec:wr140_radio}
allow prediction of the IC and relativistic bremsstrahlung flux, and
by specifying the ratio of the energy density of electrons to ions it
is possible to predict the flux from the decay of neutral pions. Such
calculations can be used to further constrain the spectral index of
the non-thermal paricles and thus the nature of their acceleration, as
demonstrated in this section.

\begin{figure*}
\psfig{figure=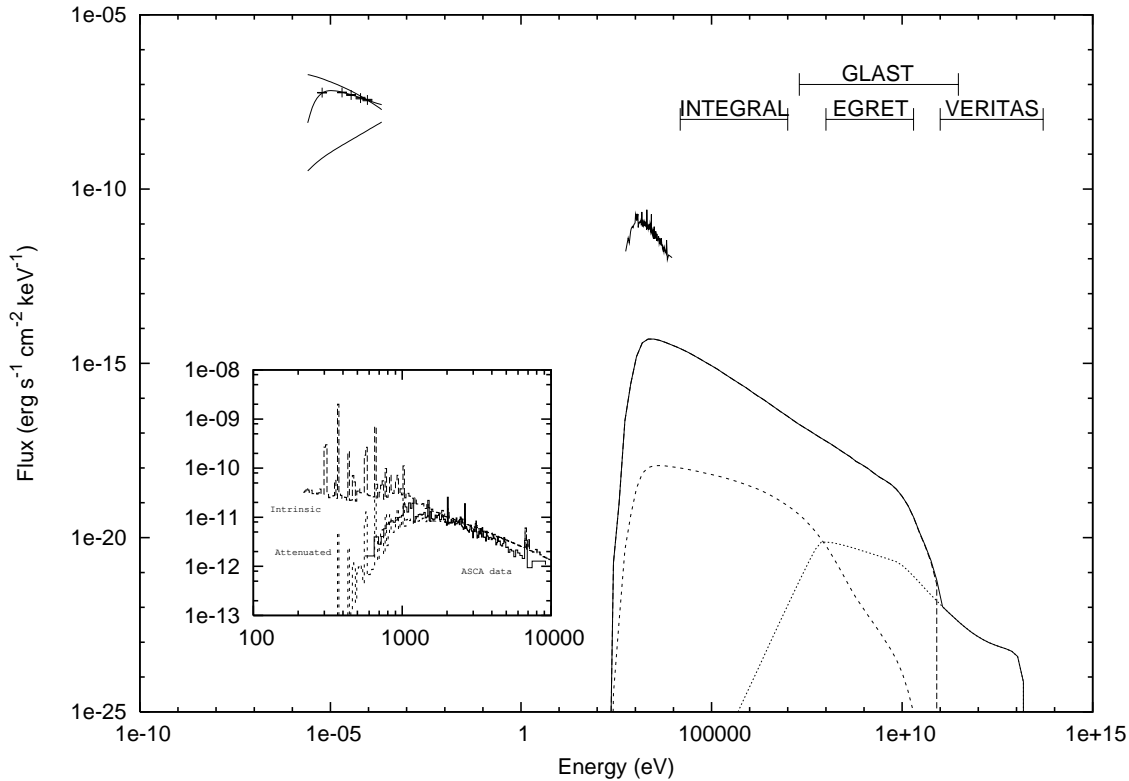,width=15.0cm}
\caption[]{The radio, and non-thermal UV, X-ray and $\gamma$-ray
emission calculated from model~B, together with the observed radio and
X-ray flux. The radio model shown indicates the thermal free-free flux
(displayed below the data points), the {\em intrinsic} synchrotron
flux before free-free absorption (displayed above the data points),
and the total observed emission (\cf~Fig.~\ref{fig:ffabs_fits}).  The
model IC (long dash), relativistic bremsstrahlung (short dash), and
pion decay (dotted) emission components are shown, along with the
total emission (solid).  Photoelectric absorption is significant at
soft X-ray energies up to $\sim 2$~keV.  To calculate the IC and
relativistic bremsstrahlung emission, $\zeta_{\rm rel,e}=5.36 \times
10^{-3}$, as derived from the fit to the radio data
(Fig.~\ref{fig:ffabs_fits}) was assumed. $\zeta_{\rm rel,i}=100 \times
\zeta_{\rm rel,e}$ was assumed to calculate the neutral pion decay
emission.  The observed {\it ASCA} X-ray spectrum from $\phi=0.837$ is
also shown. The intrinsic (dashed line) and attenuated (dotted line)
thermal X-ray emission from model~B is displayed in the inset, where
we have assumed unmodified collisional shocks, rapid thermalization of
the post-shock electrons, and collisional ionization equilibrium in
our model. The ISM absorbing column was fixed at $N_{\rm H} =
5.4\times 10^{21} \;{\rm cm}^{-2}$ \citep{Zhekov:2000}, and the
absorption coefficients for the ISM and the WR and O star winds were
calculated using Cloudy v94.00 \citep[][see
http://www.nublado.org]{vanHoof:2000}.}
\label{fig:nt_modelB}
\end{figure*}

The predicted non-thermal emission from model~B is shown in
Fig.~\ref{fig:nt_modelB}, and a summary of the emission components
noted in Table~\ref{tab:nt_lum_modelB}. The INTEGRAL flux is
insensitive to reasonable assumptions about the value of $\gamma_{\rm
max}$, though the EGRET and GLAST fluxes are more so. However, the
VERITAS flux is highly sensitive to both $\gamma_{\rm max}$ and
$\zeta_{\rm rel,i}/\zeta_{\rm rel,e}$, and the predictions presented
here are meant only as a guide. At GeV energies and above we may
underestimate slightly the flux through our neglect of the emission
from secondary particles. 

The total luminosity from non-thermal processes is $\sim 0.5$\% of the
kinetic power in the wind-wind collision ($\sim 10^{36} \ergps$), the
latter being $\sim 1$\% of the total kinetic power in the winds.  The
IC emission dominates for photon energies less than 50~GeV, while the
emission from pion decay reaches energies up to $15\;{\rm TeV}$ (the
maximum photon energy from each process is limited by the assumed
value of $\gamma_{\rm max}=10^{5}$ for the electrons and ions). The
relativistic bremsstrahlung emission is a minor contributor to the
total non-thermal flux, and does not dominate at any energy. The
absorption of high energy photons by two-photon pair production is
significant at GeV and TeV energies, in good agreement with the
calculation in Fig.~\ref{fig:gamma_tau} where a single line of sight
from the stagnation point of the WCR to an observer was considered.

The IC flux is directly proportional to the energy density of
non-thermal electrons, $U_{\rm rel,e}$, and thus to $\zeta_{\rm
rel,e}$, which in the calculation for Fig.~\ref{fig:nt_modelB} was set
to $5.36 \times 10^{-3}$ as determined from the fits of the radio data
(Fig.~\ref{fig:ffabs_fits}). The gradual steepening of the 
IC spectrum between 2-100~keV
(corresponding to $14 \ltsimm \gamma \ltsimm 100$) indicates the
effect of IC cooling on the non-thermal electron energy spectrum.  The
lack of a clearly defined spectral break reflects the situation for
the synchrotron spectrum at radio frequencies \citep[see Figs.~4 and~6
in][]{Pittard:2006}. Above 1~MeV, the spectral index is equal to
$-p/2 = -0.7$ (corresponding to a photon spectral index, 
$\Gamma = -\alpha+1 = (p+2)/2 = 1.7$).

The relativistic bremsstrahlung spectrum is fairly flat below 1~MeV,
and falls off sharply at higher energies. Like the IC emission, its
flux is directly proportional to the specified value of $\zeta_{\rm
rel,e}$.  In contrast, the emission from neutral pion decay peaks at
$0.5 {\rm m_{\pi}} c^{2} = 67.5$~MeV, and, for pions produced from
collisions involving non-thermal protons, has a maximum energy of
$E_{\rm p,max}/6$, where the maximum energy of non-thermal protons,
$E_{\rm p,max}=\gamma_{\max}{\rm m_{p}}c^{2}$.  The flux from pion-decay
is directly proportional to $n_{\rm i}U_{\rm rel,i}V$, where $n_{\rm
i}$ is the number density of thermal ions, $U_{\rm rel,i} = \zeta_{\rm
rel,i} U_{\rm th}$ is the energy density of non-thermal ions, and $V$
is the volume occupied by the WCR. Since $U_{\rm th} \propto n_{\rm i}
\propto D^{-2}$, $U_{\rm rel,i} \propto \zeta_{\rm rel,i} D^{-2}$. In
the adiabatic limit, the volume of the WCR scales as $D^{3}$, so the
intrinsic flux from neutral pion decay should scale as $\zeta_{\rm
rel,i} D^{-1}$. As we do not know how (or if) $\zeta_{\rm rel,i}$
varies with $D$, the overall dependence on $D$ is unknown. However,
since the optical depth from two-photon pair production increases as
$D^{-1}$, the observed flux above 100~GeV is likely to decline as $D$
is reduced. An added complication is that $\gamma_{\rm max}$ for
the non-thermal ions will increase with decreasing $D$ if the stars
move sufficiently close that the geometry of the magnetic field in the
unshocked stellar winds changes from toriodal to radial. In the radial
limit, we expect that $\gamma_{\rm max} \propto 1/D$, and thus
$\gamma_{\rm max}$ may reach values of $10^{6}$ in close systems
\citep[cf.][]{Bednarek:2005}.  Table~\ref{tab:nt_lum_modelB}
summarizes the emission from the model components in various
instrumental bands.

Fig.~\ref{fig:nt_modelB} also compares the non-thermal X-ray and
$\gamma$-ray emission calculated from our model with the observed ASCA
data at $\phi=0.837$ - dataset 27022010, observed on 1999-10-22.  The
O star is in front of the WR star at this orbital phase, and although
the IC emission will be slightly higher when a full treatment of the
anisotropy of the scattering is accounted for, the increase should be
less than a factor of 3 \citep[see Fig.~10 in][]{reimer:2006}. It is
reassuring that even with this factored in, the model IC fluxes are
significantly lower than the observed X-ray emission, consistent with
the lack of detection of an underlying power-law component. The inset of
Fig.~\ref{fig:nt_modelB} shows the intrinsic and attenuated thermal
X-ray emission from model~B. An important point is that the attenuated 
emission is in close agreement with the ASCA data.
Unfortunately, this is also the case for our model~A, and so
the X-ray absorption at phase 0.837 cannot be used to constrain $\eta$.

\begin{table*}
\begin{center}
\caption[]{The high energy non-thermal emission from model~B of WR\thinspace140
at orbital phase 0.837, with $\eta=0.02$ and $\zeta_{\rm rel,e}=5.36 \times
10^{-3}$. $\zeta_{\rm rel,i}=100 \times \zeta_{\rm rel,e}$ is assumed for the
calculation of the emission from neutral pion decay. 
The IC emission dominates the total emission
in each energy band noted in this table.}
\label{tab:nt_lum_modelB}
\begin{tabular}{l|cc|ccc}
\hline
\hline
Mechanism & Total emission & EGRET & EGRET & INTEGRAL IBIS & GLAST\\
          &                & (100MeV-10GeV) & (100MeV-10GeV) & (15keV-10MeV) &
(20MeV-300GeV) \\
          & (\ergps) & (\ergps) & (\phpscm2) & (\phpscm2) & (\phpscm2) \\
\hline
Inverse Compton & $3.5 \times 10^{33}$ & $2.1 \times 10^{33}$ & $5.4 \times 10^{-9}$ & $2.7 \times 10^{-6}$ & $1.5 \times 10^{-8}$ \\
Rel. bremsstrahlung & $1.8 \times 10^{30}$ & $3.2 \times 10^{29}$ & $2.5 \times 10^{-12}$ & $2.0 \times 10^{-9}$ & $2.1 \times 10^{-11}$ \\
$\pi^{0}$ decay & $6.8 \times 10^{31}$ & $9.8 \times 10^{30}$ & $1.3 \times 10^{-11}$ & $3.9 \times 10^{-14}$ & $1.8 \times 10^{-11}$ \\
\hline
\end{tabular}
\end{center}
\end{table*}

\begin{figure}
\begin{center}
\psfig{figure=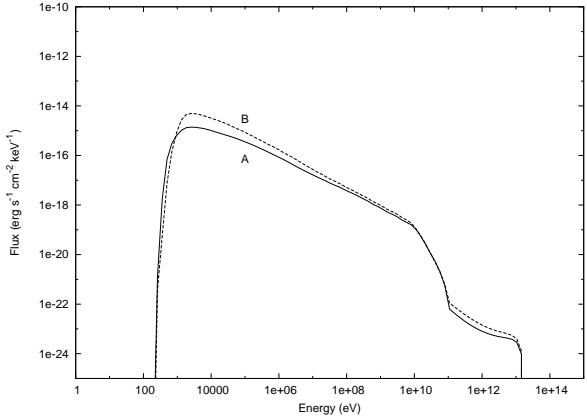,width=8.0cm}
\caption[]{The high energy non-thermal emission calculated from models
A and B with $p=1.4$.}
\label{fig:modelsAB}
\end{center}
\end{figure}

Models of the radio data alone are ill-constrained (see
Sec.~\ref{sec:wr140_radio}), and the properties of their high energy
non-thermal emission provides an important tool for discriminating between
the models. In Fig.~\ref{fig:modelsAB} , the high-energy spectra
of models~A and B, are shown. The flux from model~B is slightly higher
than from model A because $\zeta_{\rm rel,e}$ and $\zeta_{\rm rel,i}$
are slightly higher, and the IC emission has a slightly steeper
spectral slope in model B due to stronger IC cooling of the
non-thermal electron energy spectrum resulting from the smaller
distance between the apex of the WCR and the O star.

\begin{figure}
\begin{center}
\psfig{figure=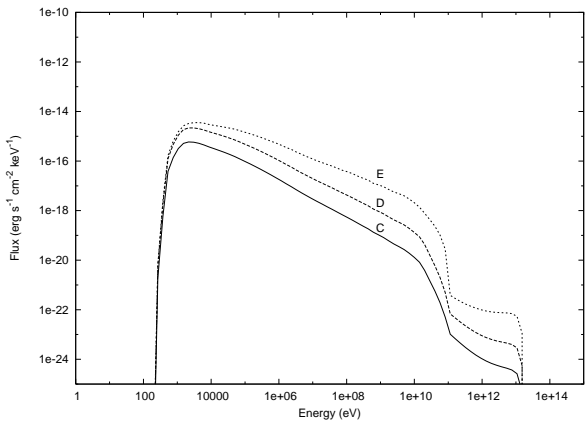,width=8.0cm}
\caption[]{The high energy non-thermal emission calculated from models
C, D and E, with $\eta=0.11$.}
\label{fig:modelsCDE}
\end{center}
\end{figure}

Fig.~\ref{fig:modelsCDE} shows the high energy non-thermal emission
from models~C, D, and E. The degeneracy of the emission at radio 
wavelengths is broken at high energies where clear differences in the
models occur. In particular, the decrease in $p$ from
model C ($p=1.53$) to model E ($p=1.1$) is clearly manifest in the
spectral slope, while the increase in $\zeta_{\rm rel,e}$
from model C to E is apparent in the relative normalization of the
spectra. Since model E has $\zeta_{\rm rel,e} = 2.8 \times 10^{-2}$
which precludes setting  $\zeta_{\rm rel,i} = 100 \zeta_{\rm rel,e}$,
we instead set $\zeta_{\rm rel,i} = 30 \zeta_{\rm rel,e}$ for this
model. 

\begin{figure}
\begin{center}
\psfig{figure=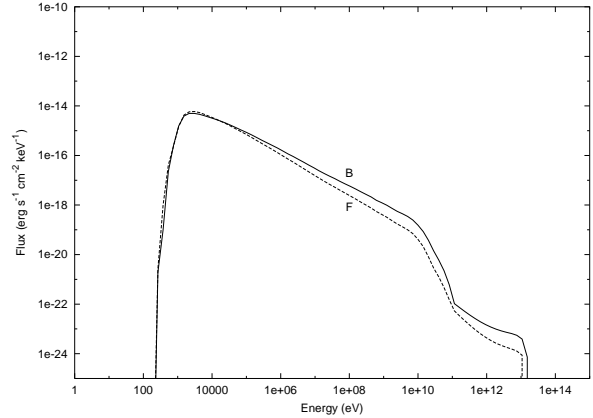,width=8.0cm}
\caption[]{The high energy non-thermal emission calculated from models
B and F, with $\eta=0.02$.}
\label{fig:modelsBF}
\end{center}
\end{figure}

The high energy non-thermal emission from models B and F are compared
in Fig.~\ref{fig:modelsBF}, where the main difference is in the assumed
stellar luminosities. The spectral slope of the IC emission at 
energies greater than 0.1~MeV reflects the different values of $p$ 
in these models. This emission arises predominantly from
those electrons near the apex of the WCR with initial post-shock
Lorentz factors greater than $\sim 100$. All such electrons 
rapidly cool to Lorentz factors of approximately 12 (model~B) and 25 (model~F)
before advecting out of the system. Although the absorption of $\gamma$-rays
by pair-production is lessened in model~F due to the reduced stellar
luminosities assumed, the commensurate reduction in $\zeta_{\rm rel,e}$
(and thus $\zeta_{\rm rel,i}$) and the increase in $p$ lead to an overall
decrease in the VERITAS flux (Table~\ref{tab:nt_lum_models}).

\begin{figure}
\begin{center}
\psfig{figure=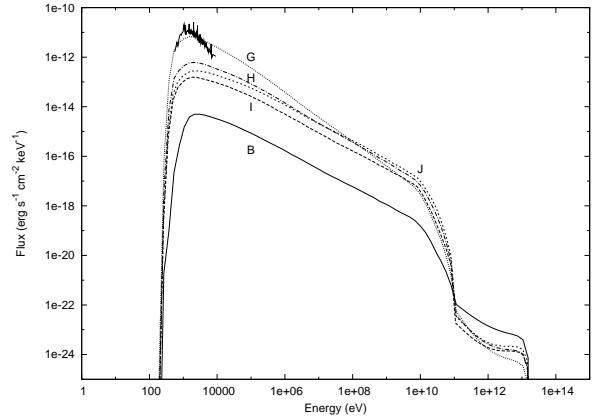,width=8.0cm}
\caption[]{The high energy non-thermal emission calculated from models
B, and G-J. The low frequency radio turnover in model~B is due to
free-free absorption, while in models G-J it is due to the Razin effect.
The {\it ASCA} data are also shown.}
\label{fig:modelsBGHIJ}
\end{center}
\end{figure}

Fig.~\ref{fig:modelsBGHIJ} shows a comparison between the high energy
non-thermal emission from models where free-free absorption (model B)
or the Razin effect (models G, H, I and J) is responsible for the low
frequency radio turnover. The large values of $\zeta_{\rm rel,e}$ in
models G-J cause the IC flux to be at least two orders of magnitude
higher relative to model B. Future observations with AGILE and GLAST
should have the sensitivity to detect such fluxes, thus allowing a
process of discrimination between our models. In fact, we can already
exclude model~G, since its predicted non-thermal emission exceeds the
observed {\em thermal} emission at X-ray energies.  While models H-J
cannot be excluded in the same manner, there are other reasons why
they are less desirable than model~B (see
Sec.~\ref{sec:radio_phi0.837}).  Table~\ref{tab:nt_lum_models} notes
the non-thermal luminosities and fluxes in various energy bands from
each model.

\subsubsection{Comparison to observations}
\label{sec:nt_comp_obs}
The total photon flux in the EGRET band from models A-F (where
free-free absorption is responsible for the low frequency radio
turnover) varies from $4.8 \times 10^{-10} \;\phpscm2$ (model~C) to $4.2
\times 10^{-8} \;\phpscm2$ (model~E). The highest prediction is still
below the flux detected from 3EG~J2022+4317 during the December 1992
observation at $\phi=0.97$ ($2.47\pm0.52 \times 10^{-7} \;\phpscm2$).  
While the different orbital
phases of the radio and EGRET data prevent a direct comparison, 
the IC emission should increase between phase 0.837 and 0.97 as
the stellar separation decreases by about a factor of three. Thus, the models
with the highest predicted fluxes are broadly consistent with the
data. In contrast, models G, H and J have predicted photon fluxes in the EGRET
band which exceed that observed from 3EG~J2022+4317. This is an additional
reason why these models are not as desirable as those where free-free
absorption is responsible for the low frequency radio turnover.

In model~B, the photon spectral index of the total non-thermal
emission, $\Gamma$, is -2.9 at 1~keV (due to photoelectric
absorption), 1.7 at 1~MeV and 100~MeV, 1.8 at 1~GeV, and 2.5 at
10~GeV, and is in rough agreement with the observed EGRET value of
$\Gamma = 2.31 \pm 0.19$ between $100\;{\rm MeV}-10\;{\rm
GeV}$. Model~E has the flattest non-thermal spectrum, with
$\Gamma=-1.0$ at 1~keV, 1.55 at 1~MeV and 100~MeV, 1.6 at 1~GeV, and
2.1 at 10~GeV, and matches the observed spectral slope less well.
 
\begin{table*}
\begin{center}
\caption[]{The high energy non-thermal emission calculated from 
the models in Table~\ref{tab:goodfits}.}
\label{tab:nt_lum_models}
\begin{tabular}{l|cc|ccccc}
\hline
\hline
Model & Total emission & EGRET & EGRET & INTEGRAL IBIS & GLAST & VERITAS \\
          &                & (100MeV-10GeV) & (100MeV-10GeV) & (15keV-10MeV) &
(20MeV-300GeV) & (200GeV-50TeV) \\
          & (\ergps) & (\ergps) & (\phpscm2) & (\phpscm2) & (\phpscm2) & (\phpscm2) \\
\hline
A & $2.4 \times 10^{33}$ & $1.5 \times 10^{33}$ & $3.4 \times 10^{-9}$ & $1.0 \times 10^{-6}$ & $1.0 \times 10^{-8}$ & $2.7 \times 10^{-14}$\\
B & $3.6 \times 10^{33}$ & $2.1 \times 10^{33}$ & $5.4 \times 10^{-9}$ & $2.7 \times 10^{-6}$ & $1.5 \times 10^{-8}$ & $4.5 \times 10^{-14}$\\
C & $3.2 \times 10^{32}$ & $1.8 \times 10^{32}$ & $4.8 \times 10^{-10}$ & $3.0 \times 10^{-7}$ & $1.4 \times 10^{-9}$ & $3.5 \times 10^{-15}$\\
D & $2.8 \times 10^{33}$ & $1.6 \times 10^{33}$ & $4.0 \times 10^{-9}$ & $1.5 \times 10^{-6}$ & $1.1 \times 10^{-8}$ & $2.8 \times 10^{-14}$\\
E & $3.6 \times 10^{34}$ & $2.0 \times 10^{34}$ & $4.2 \times 10^{-8}$ & $4.2 \times 10^{-6}$ & $9.4 \times 10^{-8}$ & $2.8 \times 10^{-13}$\\
F & $1.3 \times 10^{33}$ & $7.0 \times 10^{32}$ & $2.0 \times 10^{-9}$ & $2.4 \times 10^{-6}$ & $6.3 \times 10^{-9}$ & $1.5 \times 10^{-14}$\\
G & $2.4 \times 10^{35}$ & $7.6 \times 10^{34}$ & $2.8 \times 10^{-7}$ & $1.4 \times 10^{-3}$ & $1.1 \times 10^{-6}$ & $8.0 \times 10^{-15}$\\
H & $1.9 \times 10^{35}$ & $1.1 \times 10^{35}$ & $3.1 \times 10^{-7}$ & $2.7 \times 10^{-4}$ & $9.5 \times 10^{-7}$ & $9.9 \times 10^{-15}$\\
I & $1.1 \times 10^{35}$ & $6.8 \times 10^{34}$ & $1.7 \times 10^{-7}$ & $8.1 \times 10^{-5}$ & $4.7 \times 10^{-7}$ & $6.5 \times 10^{-15}$\\
J & $2.4 \times 10^{35}$ & $1.5 \times 10^{35}$ & $4.0 \times 10^{-7}$ & $1.7 \times 10^{-4}$ & $1.1 \times 10^{-6}$ & $1.1 \times 10^{-14}$\\
\hline
\end{tabular}
\end{center}
\end{table*}

\subsubsection{Comparison to previous work}
\label{sec:nt_comp}
The non-thermal $\gamma$-ray emission from WR\thinspace140 has
previously been estimated by \citet{Benaglia:2003} and
\citet{reimer:2006}.  Since Benaglia \& Romero's calculations are
based on a stellar separation of 10~AU (corresponding to $\phi=0.95$),
neither our estimates, their estimates, or the EGRET detections, can
be directly compared. Their estimate for the IC luminosity in the
EGRET band of $2.1 \times 10^{34} \ergps$ is typically higher than the
values from our models A-F, but lower than those from models G-J
(\cf~Table~\ref{tab:nt_lum_models}).

There are substantial differences between this work and that of
\citet{Benaglia:2003}: their estimate is obtained through the use of
Eq.~\ref{eq:icsynclum}, in which the ratio of the IC to synchrotron
emission depends strongly on the B-field, and is based on only a
single position in the WCR. They estimate that $B=200$~mG (although
noting its value is quite uncertain). In contrast, here the IC emission is
determined from models of the radio emission.  The magnetic field is a
fitted (though ill-constrained) parameter of the models, with a
maximum at the stagnation point\footnote{Since the magnetic field is
not well constrained by the radio fits, there is a large variation in
its strength between models. For instance, at the apex of the WCR,
$B=74$~mG in model~G, rising to 3.3~G in model~C, and is 1.2~G in our
preferred model (B).}, and rapidly declines in the downstream
flow. Also, in the approach presented here the B-field directly
affects the synchrotron emission, but also indirectly affects the
calculations of the high energy emission which depend on the
population and spatial distribution of non-thermal particles
i.e. $\zeta_{\rm rel,e}$ and $\zeta_{\rm rel,i}$. Finally,
\citet{Benaglia:2003} use the {\em observed} synchrotron luminosity,
rather than the intrinsic luminosity, leading to an underestimate of
the IC luminosity.

Our estimates for the relativistic bremsstrahlung luminosity are
broadly consistent with the limit of $< 10^{32} \ergps$ noted in
\citet{Benaglia:2003}, but the pion decay luminosities are wildly
discrepant with our lowest estimate being 10 orders of magnitude
higher. Details of their calculation are noted in \cite{Benaglia:2001},
where a ``cosmic ray enhancement factor'' of 10 is assumed. This
factor is the ratio between the cosmic ray energy density at the
source (in this case the WCR) and that observed at the Earth (which is
roughly equal to the average cosmic ray energy density within the
Galaxy of $\approx 1 \eV\pcm3 = 1.6 \times 10^{-12}
\erg\pcm3$). However, at the head of the WCR, the energy density of
relativistic ions in our model~B, for example, is $\approx 0.64
\erg\pcm3$. This corresponds to a cosmic ray enhancement factor of
$\sim 10^{11}$, and accounts for the large underestimate of the pion
decay emission by \cite{Benaglia:2003}.

\citet{reimer:2006} have also calculated the expected high energy
non-thermal flux from WR\thinspace140. They conclude that the change
in the IC flux with viewing angle due to anisotropic scattering is
likely to be obscured by large variations in the energy density of the
stellar radiation fields resulting from the high orbital
eccentricity. While anisotropic IC emission and Klein-Nishina effects
are clearly processes which should be included in future models, our
work shows that other uncertainties, such as the value of $p$,
currently have a greater influence on the IC emission.

\subsubsection{Predictions for forthcoming instruments}
Our predictions for the flux in the energy band of the IBIS instrument
onboard INTEGRAL and the GLAST large area telescope (LAT) are also
noted in Table~\ref{tab:nt_lum_models}. For models A-F the predicted
fluxes are $\sim 10^{3} \times$ lower in the INTEGRAL band, and $\sim
10^{2} \times$ lower in the GLAST band than those made by
\cite{Benaglia:2003}.
Using the INTEGRAL Observing Time
Estimator\footnote{http://integral.esac.esa.int/isoc/operations/html/OTE.html},
we find that model~A cannot be detected at the $5\sigma$ level in the
$15\;{\rm keV} - 10\;{\rm MeV}$ range using the IBIS imager, since an
exposure time of 300~yrs is required! Model~E, the brightest of models
A-F in this band, still requires an exposure of 180~yrs. With several
bright sources (including Cyg X-1) nearby, the required exposure time
may be even longer. On the other hand, for model~J an exposure time of
only 2.8~Msec is needed. At the time of writing, WR\thinspace140 is
not detected in a 2~Msec exposure \citep{DeBecker:2005b}, and we are
close to ruling out model~J (and~H) on this basis. Line emission at
$\gamma$-ray energies from the de-excitation of nuclear isomers, such
as $^{12}{\rm C}^{*}$ and $^{16}{\rm O}^{*}$, is much too faint to be
detected with current instrumentation \citep[][]{Benaglia:2003}.

The GLAST $5\sigma$ sensitivity at $E > 100$~MeV for sources at high
galactic latitudes with a $\Gamma=2$ spectrum after a 2~yr all-sky
survey is $1.6 \times 10^{-9} \;\phpscm2$. Therefore, our predictions 
suggest that WR\thinspace140 will be detectable, this being
important to differentiate between the models in
Table~\ref{tab:nt_lum_models}.

At energies of approximately 50~GeV and higher, the flux from the
models arises exclusively from the decay of neutral pions, with a high
energy cut off of $\approx 15$~TeV, strongly dependent on $\gamma_{\rm max}$ 
of the ions. The VERITAS-4 array is sensitive to
$\gamma$-rays in the 200~GeV to 50~TeV energy band, but unfortunately
the high absorption from two-photon pair production severely reduces
the photon flux above $\sim 10$~GeV. The predicted flux from model~B
is not sufficient for a detection in a 50~hr observation 
\citep[cf.][]{Fegan:2003}, and we
conclude that WR\thinspace140 is probably too faint to be detected at
$\phi=0.837$, though it might be brighter at phases closer to
periastron. An observation by GLAST, for example, will better
constrain the potential models, and allow more robust predictions
of the flux at $\sim$~TeV energies.

\section{Discussion}
\label{sec:discuss}
\subsection{Mass-loss rates of early-type stars}
It is important that accurate measurements of mass-loss rates from early-type 
stars are obtained, as mass-loss is known to significantly affect their
evolution, and the injection of mass, momentum and energy influences
not only the gas-phase conditions in the immediate environment of the
clusters within which the massive stars form, but also the phase
structure and energetics of the ISM on galactic scales, and the
thermodynamics and enrichment of the intergalactic medium on Mpc
scales and above.

Unfortunately, observational determinations of mass-loss rates are
often highly uncertain, in part because many of the techniques are
sensitive to the degree of structure in the winds. Overestimates by
factors of at least 3 or more are likely when the winds are clumpy,
but for some types of O stars it could be as high as factors of 20 or
more \citep{Fullerton:2006}. The X-ray emission from CWB systems
provides an opportunity to measure mass-loss rates using a method
which is not expected to be very sensitive to clumping (see
Sec.~\ref{sec:xray_mdots}).  In fact, mass-loss rates could be
underestimated using this technique: at least some winds may be both
denser and faster at high latitudes
\citep{Dwarkadas:2002,Smith:2002,Chesneau:2005}, and if the stellar
rotation axes of the stars are closely perpendicular to the orbital
plane, the WCR then probes those parts of the winds which are less
dense and slower. In addition, a significant amount of energy may be
siphoned into particle acceleration.  In this work, the mass-loss rate
for the O supergiant star in WR\thinspace140 is towards the lower end
of previous estimates, but is consistent with expected biases. A
reduction in mass-loss rates, and/or mass-loss through a predominantly
polar flow, can also explain the near-symmetry of observed X-ray lines
from early-type stars
\citep{Owocki:2006,Mullan:2006,Cohen:2006}, and
both reduced mass-loss and polar flow help in the theoretical
formation of gamma-ray bursts \citep{Woosley:2006}.

\subsection{The nature of shocks and particle acceleration in wide CWB systems}
\label{sec:nature}
In this work we have presented evidence for a fairly hard spectrum
(i.e. $p<2$) of non-thermal electrons in WR\thinspace140, which is
further supported by earlier fits to the radio data of WR\thinspace147
\citep[this was mistakenly interpreted as evidence for shock
modification in][]{Pittard:2006}. As discussed in
Sec.~\ref{sec:dsa}, there are many potential mechanisms for producing
$p<2$. One possibility is particle re-acceleration, but while the
wind-embedded shocks believed to occur in stellar winds due to the
line de-shadowing instability of the radiative driving
\citep[e.g.,][]{Owocki:1988} may provide such an opportunity, and thus
harden the non-thermal particle spectrum prior to the global shocks
bounding the WCR, further evidence is needed for this explanation to
be more persuasive, not least because re-acceleration has yet to be
demonstrated in theoretical calculations of wind-embedded shocks. One
problem is that in current one-dimensional simulations of
wind-embedded shocks, the accelerated particles are trapped between
forward and reverse shocks. This has posed a problem for the
interpretation of non-thermal radio emission from single stars
\citep{vanLoo:2006}, and only calculations using a
phenomenological shock model are currently successful in fitting the
data \cite[e.g.,][]{Chen:1994}. Two- and
three-dimensional models of instability-generated structure
\citep[see, e.g.,][]{Dessart:2003} may allow accelerated particles to
escape one shock and be re-accelerated at others, but this has still
to be demonstrated.

We can speculate about additional effects which occur in CWB systems
with highly eccentric orbits. For example, if re-acceleration is
important, one might expect $p$ to depend on the orbital phase, as the
number of wind-embedded shocks that accelerate a non-thermal particle
prior to the WCR may change significantly as the stellar separation
alters.  Similarly, if the obliquity of the shocks changes with
stellar separation, the injection efficiency and rate of particle
acceleration may also alter. In addition, it is likely that the
maximum energy obtained by electrons and ions varies with stellar
separation. Finally, the assumption of isotropic synchrotron emission
will need to be reconsidered if the magnetic turbulence is low enough
that the mean B-field has a preferred direction.  In the context of
WR\thinspace140, these are all potential explanations for the observed
asymmetry of the radio lightcurve.

Is the efficiency of particle acceleration in WR\thinspace140
high enough for shock modification to occur? In Tycho's SNR, the
concave curvature of synchrotron radio emission is clear, though still
indirect, evidence for shock modification
\citep[see][]{Volk:2002}. Unfortunately, in WR\thinspace140 the
observed synchrotron emission extends over a much narrower frequency
range and a concave curvature cannot easily be distinguished. And
while the lower than expected thermal X-ray temperature is consistent
with shock modification, a more probable cause is $T_{\rm e} < T_{\rm
i}$. Likewise, while shock modification may help to explain the
behaviour of the X-ray lightcurve which does not follow the expected
$1/D$ variation for an adiabatic WCR \citep{Pollock:2002}, there is a
realistic chance that radiative cooling is the cause.  Finally, the
B-field in WR\thinspace140 is sufficiently high to cause significant
energy to be transferred from the non-thermal particles into magnetic
turbulence via Alfv\'{e}n wave heating, thereby lowering the
acceleration efficiency, although it remains possible that shock
modification occurs predominantly through particle loss (due to the
curvature of the shocks) rather than high acceleration
efficiencies. To summarize, there is currently no strong evidence for
shock modification in WR\thinspace140, and on balance we conclude that
strongly modified shocks are unlikely to occur in WR\thinspace140.

\subsection{Non-thermal X-ray, TeV and muon neutrino emission from CWBs}
While there is no observational evidence for non-thermal X-ray
emission in WR\thinspace140, there are some systems that appear to
show a power-law tail.  Perhaps the strongest candidate is
$\eta$~Carinae, where X-rays with a photon index $\Lambda\approx2.4$ were
detected at energies of up to 150~keV during a long BeppoSAX exposure 
\citep{Viotti:2004}.
Determining the spectal slope at higher energies should allow
discrimination of the competing models proposed to explain this
emission.

The lack of observed non-thermal X-ray emission from CWB systems which
are known non-thermal radio emitters
\citep{DeBecker:2004,DeBecker:2005,DeBecker:2006} contrasts with the
possible detection of non-thermal X-ray emission from the 3.4~d
orbital period system HD~159176, which is not known to be a
non-thermal radio emitter \citep{DeBecker:2004b}.  Though the nature
of the wind-wind collision in such short period systems is uncertain,
if particle acceleration occurs, perhaps mainly through reconnection,
IC losses will be catastrophic, and it is not surprising that
non-thermal X-ray emission is possibly detected despite no clear
signature of synchrotron emission \citep[due to the obvious high
absorption - ][]{DeBecker:2005b}.

%

In the models of WR\thinspace140 presented in this paper, absorption
of high energy photons by pair-production creates a very steep photon
spectrum between 10~GeV and 1~TeV. Consequently, the photon flux above
1~TeV is many times lower than the flux predicted in the VERITAS
bandpass, which is dominated by photons with energies less than 1~TeV
(see Table~\ref{tab:nt_lum_models}). For example, in model~B the flux
above 1~TeV is $1.2 \times 10^{-14}\;\phpscm2$. Nevertheless, the flux
from CWB systems compares favourably to some other mechanisms proposed
for the generation of TeV emission, such as the interaction of cosmic
rays in the innermost parts of the winds of early-type stars
\citep{Torres:2004} or within the cluster wind of a dense stellar
cluster \citep{Domingo:2006}. Therefore, it is possible that several
CWBs or multiple colliding winds from a cluster of stars contribute
part or all of the flux from the unidentified TeV source detected in
the Cygnus region, TeV~J2032+4130 \citep{Aharonian:2005a}, which has
an integral flux $F_{\gamma}(E_{\gamma} > 1\;{\rm TeV}) = 6.9(\pm1.8)
\times 10^{-13} \;\phpscm2$. 
The unidentified TeV source HESS~J1813-178 (with an
integrated flux above 200~GeV of $1.2\times10^{-12}\;\phpscm2$) may
also be associated with an OB association \citep{Aharonian:2005b}.

\begin{figure}
\begin{center}
\psfig{figure=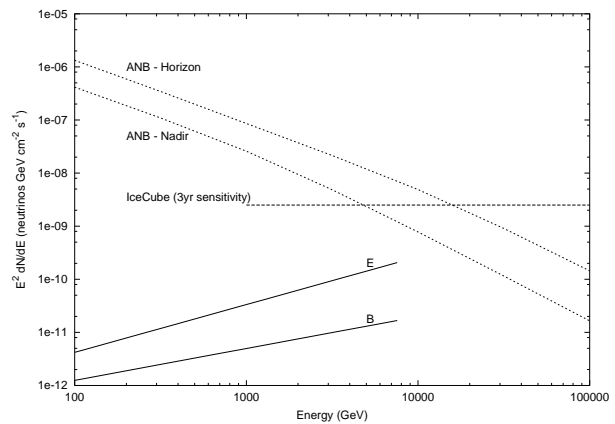,width=8.0cm}
\caption[]{The differential spectra of muon neutrinos produced by
neutral pion decay from models B and E (the fluxes have been reduced by
a factor of 2 to account for neutrino oscillation). The atmospheric neutrino
background flux within $1^{\circ}$ of the source \citep{Lipari:1993} 
is shown, together with the 3-yr sensitivity of the IceCube detector
\citep{Hill:2001}.}
\label{fig:neutrino}
\end{center}
\end{figure}

The decay of charged pions created in hadronic collisions between
non-thermal and thermal ions produces a $\nu_{\mu}+\bar{\nu}_{\mu}$ 
neutrino flux. This flux can be estimated from the $\gamma$-ray flux 
produced by neutral pion-decay by imposing energy conservation
\citep[][]{Alvarez:2002}.  Fig.~\ref{fig:neutrino} shows the
resulting flux from models~B and E, including neutrino oscillation. 
It is clear that the predicted fluxes are neither above the
atmospheric neutrino background (ANB), nor above the sensitivity limit
of the proposed IceCube detector.
Thus, our finding is contrary to the conclusion of
\citet{Bednarek:2005}, who predicted a detectable muon neutrino flux from
WR~20a. This discrepancy is due to Bednarek's use of a higher 
conversion efficiency of kinetic power into non-thermal
particles, largely through the assumption $\eta=1$.

\subsection{Colliding winds in dense stellar clusters}
Our work is also relevant to more complicated systems
involving colliding winds, such as dense stellar clusters.
Hydrodynamical simulations of the multiple colliding winds in clusters
have been performed by \citet{Raga:2001}, with particular application
to the Galactic Center in \citet{Coker:1997}, \citet{Coker:2005},
\citet{Rockefeller:2005}, and \citet{Cuadra:2006}. A network of
bubbles forms around the stars, which are embedded in a medium of hot
shocked gas. Highly non-linear DSA should occur at the stellar
termination shocks in stellar clusters if the pre-shock B-field is
small enough \citep[$B \ltsimm 10 \;\mu{\rm G}$,][]{Ellison:2005b},
though the most extreme modification (e.g., compression ratios much 
greater than 4) will occur only at the cluster's edge.

Shock modification is also a potential explanation for the lower than
expected temperature of the X-ray emitting gas observed in several
dense stellar clusters \citep[e.g.,][]{Stevens:2003}, though
mass-loading and slow equilibration of the post-shock ion and electron
temperatures are other possibilities. In addition, the X-ray
luminosities of the diffuse gas are generally higher than expected,
which indicates mass-loading and/or a low thermalization efficiency
\citep{Stevens:2003,Oskinova:2005}, the latter also being consistent
with efficient particle acceleration and shock modification.  If the
rate at which energy is removed from the cluster (for example, by
radiative cooling) exceeds 30\% of the energy injection rate, a time
independent cluster wind cannot form \citep{Silich:2004}. Though the
removal of energy through particle acceleration was not considered,
efficiencies of this order are expected \citep[e.g.,][]{Bykov:1999},
in which case a full time-dependent solution for the cluster wind is
required.

Amongst the dense stellar clusters known in our Galaxy, the central
cluster around Sgr~A* is unusual in many respects, including the
recent detection of apparently thermal gas with a temperature $kT
\approx 8$~keV \citep{Muno:2004}, and TeV emission. Observations with
HESS show a 1-10 TeV luminosity of $\sim10^{35}\ergps$
\citep{Aharonian:2004}. The upper limit on the TeV source size is $<3'$ 
(95\% confidence level), which corresponds to $<7$~pc at the
distance of the Galactic Center.  The photon spectrum above the
165~GeV threshold is specified by $F(E) = F_{\rm 0} E_{\rm
TeV}^{-\Gamma}$, with $\Gamma\approx2.2$ and $F_{\rm
0}\approx2.5\times10^{-12}\;\phpscm2\;{\rm TeV}^{-1}$, and has been
independently confirmed with MAGIC \citep{Albert:2006}.
Various origins of the TeV emission have been proposed, including the
central supermassive black hole \citep{AharonianN:2005a}, the young
SNR Sgr~A East \citep{Crocker:2005}, the dark matter
halo \citep{Horns:2005}, and from non-thermal particles accelerated at 
the multiple colliding winds shocks \citep{Quataert:2005}, which we
now discuss.

The total wind power estimated from the cluster of stars within
$\sim10''$ ($\sim0.4$~pc) of the Galactic Center is $\sim3 \times
10^{38} \ergps$.  In the context of colliding winds, the TeV emission
may arise from IC upscattering of ambient photons by non-thermal
electrons, or from the decay of neutral pions created by the
collisions of non-thermal and thermal hadrons. We now consider each
process in turn. It is likely that IC cooling limits the maximum
Lorentz factor that non-thermal electrons within the central parsec
can attain to $\sim10^{7}$ \citep{Quataert:2005}, which is only
marginally high enough for IC scattering to account for the detected
TeV flux.  The TeV emission occurs from electrons whose cooling time
is less than the flow time out of the cluster i.e. from those with
$\gamma > \gamma_{\rm c}$. For a power-law distribution of electrons
$n(\gamma) \propto \gamma^{-p}$, the IC photon spectral index
is $\Gamma=(p+2)/2$, for $\gamma > \gamma_{\rm c}$.  Hence the TeV
spectral index of 2.2 implies that $p\approx2.4$.  If instead neutral
pions are responsible, the observed spectral index indicates that
$p\approx2.2$.  In the context of multiple colliding winds,
\citet{Ozernoy:1997} estimates the emission from pion-decay to be
$\sim 3 \times 10^{36}\ergps$, assuming 10\% acceleration efficiency
for hadrons at the multiple shocks, and a 10\% energy transfer into
pions i.e. a 1\% efficiency overall. Other estimates in the literature
suggest a somewhat lower luminosity from this process
\citep[e.g.,][]{Quataert:2005}, but this process nevertheless
appears to be plausible on energy grounds.

The required value of $p$ for each of these processes is much softer
than we find here for WR\thinspace140, and also for the theoretical limit for
multiple shocks \citep[p=1.0,][]{Pope:1994}. This implies that
re-acceleration at multiple shocks within the cluster, as proposed by
\citet{Ozernoy:1997}, does not occur.
It is possible to imagine several reasons why this might be the case.
Firstly, there is unlikely to be a seed population of non-thermal
particles from wind-embedded shocks encountering the termination shock
of the winds because of the large adiabatic cooling experienced as the
plasma flows out to $\sim 0.01-0.1$~pc. Hence, the particle spectrum
immediately downstream of the termination shocks is likely to
correspond to the single-shock limit.  Secondly, it is not entirely
clear that there will be numerous opportunities for particle
re-acceleration within the radius of the cluster, since the cooling
timescale of the shocked gas is long, and it remains hot, and
therefore subsonic. Analytical solutions of cluster winds reveal that
the shocked gas, on average, only becomes supersonic at the outer
radius of the cluster \citep{Chevalier:1985}. Within the cluster
interior, the network of hot post-shock gas
experiences pressure gradients and weak shocks as it escapes, but
in general is not likely to be repeatedly shocked.
This lack of opportunity for re-acceleration means that $p$ is likely
to remain close to the value obtained from a single shock. 

Perhaps the most
natural explanation is that the majority of particle acceleration occurs
at low Mach number shocks ($M\approx3$ and $M\approx5$, for IC or 
pion-decay emission respectively) within the network
of hot shocked gas. This conclusion does not rule out the possibility
of modification at the termination shocks of the individual winds,
which would be consistent with the low X-ray temperature of the cooler
emission component \citep[$kT\approx0.8$~keV,][]{Muno:2004}, 
but does exclude shock modification as the only process since a
hard particle spectrum ($p<2$) is expected at the high particle
energies ($\gamma \gtsimm 10^{5}$) which are necessary. An alternative
explanation is sub-diffusive cross-field transport (see Sec.~\ref{sec:dsa}).

The Galactic Center region has also been detected at GeV energies.
The observed luminosity of the EGRET source 3EG~J1745-2852 is $\approx
2\times10^{37}\ergps$ \citep{Mayer-Hasselwander:1998,Hartman:1999}.
However, while its position is consistent with Sgr~A*, the angular
resolution of $1^{\circ}$ covers a solid angle about 100 times larger
than the emission seen with HESS. As the EGRET spectrum is about 4
orders of magnitude brighter than the observed HESS emission
extrapolated as a power-law to GeV energies, different
processes/sources may well be responsible for the detected emission
\citep[although different assumptions concerning the injection
spectrum and diffusion coefficients of non-thermal protons may account
for both observations, as shown by][]{AharonianN:2005b}. Regardless of
the origin of non-thermal protons, if the emission is from pion-decay,
a substantial TeV neutrino flux should be seen by neutrino telescopes.

While it remains difficult to draw firm conclusions on the nature of
the TeV emission observed from the Galactic Center, we note that the
nearby Arches cluster, which has an integrated mass-loss rate from its
stellar components within an order of magnitude of the Galactic Center
cluster, has not yet been detected at TeV energies. This may argue
against the cluster wind scenario as an explanation for the TeV
emission from the Galactic Center.





\section{Summary and future directions}
\label{sec:summary}
In this paper we have applied models to the observed emission from
WR\thinspace140 at $\phi=0.837$, when the radio emission is near
maximum, and when X-ray data are available. Predictions for the
X-ray and $\gamma$-ray emission from IC scattering, relativistic
bremsstrahlung, and neutral pion decay, as well as for the muon
neutrino flux, are made. Our calculations are based on the orbital
solution of \citet{Dougherty:2005}, and contain a number of
simplifying assumptions, including i) that the non-thermal electrons
and ions are accelerated at the shocks bounding the WCR (this is
assumed to be through DSA, but need not necessarily be so) and have
power-law spectra with a sharp cutoff at $\gamma_{\rm max}=10^{5}$
before advecting downstream, ii) the non-thermal particle distribution
at these shocks is normalized with respect to the local post-shock
thermal energy density, the non-thermal ions have $100\times$ the
energy density of the non-thermal electrons, and neither the spectral
index of the energy distribution nor $\gamma_{\rm max}$ vary along the
shocks, iii) the energy density of the magnetic field is similarly
normalized to the local post-shock thermal energy density, iv)
particle acceleration has no effect on the thermal shock structure
i.e. there is no shock ``modification'', v) IC cooling occurs in the
Thomson limit with isotropic scattering, vi) the emission from
secondary particles is ignored. Rapid IC cooling of the high energy
electrons in the post-shock flow means that the local value of
$\gamma_{\rm max}$ is typically $<< 10^{5}$. The assumptions
$\gamma_{\rm max}=10^{5}$ and $\zeta_{\rm rel,i}=100\zeta_{\rm
rel,e}$, together with the assumptions in v), mean that the predicted
emission at GeV energies and above is relatively uncertain at this
stage.  Our predictions for the VERITAS-4 flux, in particular, should
be treated as an estimate.

Despite the large number of assumptions listed above, it is possible
to make a number of key deductions concerning the nature of
WR\thinspace140 and the particle acceleration process at the WCR.  We
clearly show that VLBA imaging observations of WR\thinspace140 do not
currently have enough sensitivity to constrain the wind momentum
ratio $\eta$, through determination of the shock opening angle. VLBI 
observations are only sensitive to high-brightness emission that
arises in the region of the apex of the WCR, where the WCR has yet
to attain an asympototic opening angle.
Furthermore, we find that the relative positions of the stars and the
peak of the observed emission from the WCR is not necessarily related
to $\eta$, due to occultation by the stellar winds. Such observations
have been used to determine $\eta$ in a number of wider CWB systems
where occulatation effects are expected to be relatively unimportant
\citep[e.g.][]{Dougherty:1996,
Dougherty:2000,Contreras:1997,Williams:1997}, but we caution against
the application of this method to closer systems like WR\thinspace140.
In spite of these limitations in interpreting VLBA observations, it is
possible to constrain the wind momentum ratio using simultaneous fits
to the radio and X-ray data to $\eta \approx 0.02$. A remaining
concern is that the line-of-sight angle into the system in our
preferred model is some way from the observationally determined value,
although the uncertainty in the latter is not negligible.  Additional
data is required to further constrain the orbital parameters of
WR\thinspace140.

The stellar mass-loss rates in our preferred model are $\Mdot_{\rm WR}
\approx 4.3 \times 10^{-5} \;\Msolpyr$ and $\Mdot_{\rm O} \approx 8
\times 10^{-7} \;\Msolpyr$. While the O star mass-loss rate is
substantially lower than the values estimated by \citet{Repolust:2004}
for O4-5 supergiants, an order of magnitude reduction in the mass-loss
rate is consistent with the recent estimates of
\citet{Fullerton:2006}, and with some current explanations of the
observed near-symmetry of X-ray lines. Such sizeable reductions have
extensive implications for massive star evolution (including gamma-ray
bursts), and feedback on the interstellar medium.  There is also some
scope for reducing the stellar luminosities adopted from
\citet{Dougherty:2005}.

An important finding is that models of the radio data require
$p<2$. Several possible mechanisms can in principle explain this.
In any case, a flatter spectrum for the non-thermal particles has
important implications for the modelling of the high energy
non-thermal emission from such systems, with almost all previous
models in the literature assuming that $p=2$. While our fits are
currently non-unique, models where the Razin effect is responsible for
the low frequency radio turnover (which allow $p=2$) can be ruled out
on the basis that the implied efficiency of electron acceleration is
extraordinarily high. Furthermore, they predict the detection of IC
emission by INTEGRAL which has not yet been achieved, and in some
cases non-thermal X-ray emission at fluxes higher than the observed
thermal emission. Observations with Suzaku should be more sensitive
than those made with INTEGRAL, and will allow tighter constraints
to be placed on the strength of the magnetic field. Our inability to
obtain fits to the radio data with $p>2$ does not necessarily exclude
the possibility of shock modification, but on balance we believe that
this is unlikely to occur in WR\thinspace140.  However, large
non-linear effects, comparable to those possible at the reverse shock
in SNRs \citep{Ellison:2005b}, may occur at the wind-wind collisions
in stellar clusters where the B-field is much lower.


Our flux predictions at orbital phase 0.837 are summarized in
Table~\ref{tab:nt_lum_models}. The photon fluxes in the INTEGRAL,
GLAST, and VERITAS-4 band-passes range from
$3\times10^{-7}-3\times10^{-4}$, $10^{-9}-10^{-6}$, and
$4\times10^{-15}-3\times10^{-13}$, respectively, with our preferred
model having fluxes of $3 \times 10^{-6}$, $2 \times 10^{-8}$, and $5
\times 10^{-14}\;\phpscm2$.  The high energy non-thermal emission
from WR\thinspace140 is expected to display significant variability
with orbital phase as the stellar separation and line-of-sight
angle change, both of which affect the opacity of absorption in the
stellar radiation fields and the emission from the resulting IC
cascade of secondary electrons and positrons.  In addition, the
spectral index and the acceleration efficiency of the primary
non-thermal particles may also change.  We emphasize that the future
detection of TeV emission from CWB systems will almost certainly
indicate pion-decay, since the high photon fluxes prevent the
acceleration of electrons beyond $\gamma \sim 10^{5}-10^{6}$.
WR\thinspace140, along with WR\thinspace146 which has the brightest
non-thermal radio emission of any CWB \citep{Dougherty:2000}, and
WR\thinspace147 which is the widest CWB system on the sky
\citep{Williams:1997} and the only system resolved in X-rays
\citep{Pittard:2002c}, present a real opportunity for the unambiguous
detection of pion-decay from accelerated ions. The estimated flux of
TeV photons from CWB systems compares favourably to some other
mechanisms, and may contribute part or all of the flux from some TeV
sources. The multiple colliding winds in the Galactic Center cluster
may make a non-negligble contribution to the observed TeV flux.
However, wide CWB systems are not expected to be strong neutrino
sources.

Tighter constraints on $p$ and the nature of the shocks and particle
acceleration in colliding winds systems (binaries and clusters) are
best achieved through the dual strategy of improving the theoretical
models and obtaining observational data at MeV and GeV energies. On
the theoretical side, models should explicitly account for the transfer
of part of the available pre-shock energy into accelerated particles,
self-consistently calculate the non-thermal electron and ion spectra
and their subsequent evolution as they advect with the downstream
flow, and include Klein-Nishina effects and anistropic IC
scattering. Upcoming observations by AGILE and GLAST at MeV and GeV
energies, where the emission is no longer degenerate with $B$ and $p$,
should then distinguish between the non-unique model fits to the radio
data, and determine the B-field in the WCR and the spectral index of
the non-thermal particles.

\section*{acknowledgements}
We have had many interesting discussions during the course of this
work, and would like to thank Felix Aharonian, Paula Benaglia, Andrei
Bykov, Paul Crowther, Chuck Dermer, Jamie Holder, Jim MacDonald, Anita
Reimer, Sven van Loo, and Perry Williams. We would also like to thank
Mike Corcoran for supplying the ASCA X-ray spectrum of
WR\thinspace140, Rob Coker for comments on an earlier draft, and Evan
O'Connor for running some early models. Johannes Knapp, Andy Pollock,
Ian Stevens, and Perry Williams provided helpful comments prior to
submission. Special thanks go to Don Ellison for his tireless response
to questions over the course of the last few years. We would like to
thank the referee for a constructive, timely, and helpful report.  SMD
gratefully acknowledges partial funding for a visit to Leeds. JMP is
supported by a University Research Fellowship from the Royal
Society. This research has made use of NASA's Astrophysics Data System
Abstract Service.

\label{lastpage}

\end{document}